\documentstyle[11pt]{article}

\begin{document}

\newcommand{\beq}{\begin{equation}}
\newcommand{\eeq}{\end{equation}}
\newcommand{\bea}{\begin{eqnarray}}
\newcommand{\eea}{\end{eqnarray}}

\newcommand{\chii}{\raise.5ex\hbox{$\chi$}}
\newcommand{\R}{I \! \! R}
\newcommand{\N}{I \! \! N}
\newcommand{\C}{I \! \! \! \! C}

\newcommand{\noi}{\noindent}
\newcommand{\vs}{\vspace{5mm}}
\newcommand{\ie}{{${ i.e.\ }$}}
\newcommand{\eg}{{${ e.g.\ }$}}
\newcommand{\ea}{{${ et~al.\ }$}}
\newcommand{\hf}{{\scriptstyle{1 \over 2}}}
\newcommand{\ih}{{\scriptstyle{i \over \hbar}}}
\newcommand{\hi}{{\scriptstyle{ \hbar \over i}}}
\newcommand{\itwoh}{{\scriptstyle{i \over {2\hbar}}}}
\newcommand{\dbrst}{\delta_{BRST}}

\newcommand{\deder}[1]{{ 
 {\stackrel{\raise.1ex\hbox{$\leftarrow$}}{\delta^r}   } 
\over {   \delta {#1}}  }}
\newcommand{\dedel}[1]{{ 
 {\stackrel{\lower.3ex \hbox{$\rightarrow$}}{\delta^l}   }
 \over {   \delta {#1}}  }}

\newcommand{\papar}[1]{{ 
 {\stackrel{\raise.1ex\hbox{$\leftarrow$}}{\partial^r}   } 
\over {   \partial {#1}}  }}
\newcommand{\papal}[1]{{ 
 {\stackrel{\lower.3ex \hbox{$\rightarrow$}}{\partial^l}   }
 \over {   \partial {#1}}  }}

\newcommand{\ddr}[1]{{ 
 {\stackrel{\raise.1ex\hbox{$\leftarrow$}}{\delta^r}   } 
\over {   \delta {#1}}  }}
\newcommand{\ddl}[1]{{ 
 {\stackrel{\lower.3ex \hbox{$\rightarrow$}}{\delta^l}   }
 \over {   \delta {#1}}  }}
\newcommand{\dd}[1]{{  {\delta} \over {\delta {#1}}  }}
\newcommand{\pa}{\partial}
\newcommand{\sokkel}[1]{\!  {\lower 1.5ex \hbox{${\scriptstyle {#1}}$}}}  
\newcommand{\larrow}[1]{\stackrel{\rightarrow}{#1}}
\newcommand{\rarrow}[1]{\stackrel{\leftarrow}{#1}}
\newcommand{\twobytwo}[4]{\left[\begin{array}{ccc}{#1}&&{#2} \cr
                                  {#3} && {#4} \end{array} \right]}

\newcommand{\nnn}{n}
\newcommand{\Gamm}{z}
\newcommand{\www}{w}
\newcommand{\AAA}{A}
\newcommand{\aaa}{a}
\newcommand{\FF}{{\cal F}}
\newcommand{\RR}{{\cal R}}
\newcommand{\fo}{f^{(0)}}
\newcommand{\Eta}{\eta_{(0)}}
\newcommand{\yy}{\chi}

\newcommand{\eq}[1]{{(\ref{#1})}}
\newcommand{\mb}[1]{{\mbox{${#1}$}}}

\newcommand{\proofbox}{\begin{flushright}
${\,\lower0.9pt\vbox{\hrule \hbox{\vrule
height 0.2 cm \hskip 0.2 cm \vrule height 0.2 cm}\hrule}\,}$
\end{flushright}}

\begin{titlepage}
\title{\Large{\bf Almost Parity Structure,\\
Connections and\\ 
Vielbeins in BV Geometry}}

\author{{\sc K.~Bering}
\footnote{Email address: {\tt bering@nbivms.nbi.dk}}
\\
Center for Theoretical Physics \\
Laboratory for Nuclear Science \\
and Department of Physics \\
Massachusetts Institute of Technology \\
Cambridge, Massachusetts 02139 \\
{~}}

\date{MIT-CTP-2682,~~~~~ physics/9711010. {~~~~~}November 1997 }
\maketitle
\begin{abstract}
We observe that an anti-symplectic manifold locally always admits a 
parity structure. The parity structure can be viewed as a complex-like 
structure on the manifold. This induces an odd metric and its Levi-Civita 
connection, and thereby a new notion of an odd K\"{a}hler geometry. 
Oversimplified, just to capture the idea, the bosonic variables are 
``holomorphic'', while the fermionic variables are ``anti-holomorphic''. 
We find that an odd K\"{a}hler manifold in this new ``complex'' sense 
has a nilpotent odd Laplacian iff it is Ricci-form-flat. The local 
cohomology of the odd Laplacian is derived. An odd Calabi-Yau manifold has 
locally a canonical volume form. We suggest that an odd Calabi-Yau manifold 
is the natural geometric notion to appear in covariant BV-quantization. 
Finally, we give a vielbein formulation of anti-symplectic manifolds.
\end{abstract}
\end{titlepage}

\vfill
\newpage

\setcounter{equation}{0}
\section{Introduction}

\noi
Odd symplectic geometry was first introduced in $1981$ by Batalin and 
Vilkovisky \cite{bv} in antifield formulation of gauge theories. It was 
soon realized that many of the constructions in even symplectic geometry,
could be transfered into odd symplectic geometry. However, there are 
important exceptions: For instance, there is no canonical volume element 
in the odd case. The main applications of odd symplectic geometry has up 
to now been quantization of gauge theories and covariant formulations of
string field theory \cite{bz}. Many authors Khurdaverdian and 
Nersessian \cite{khuner}, Batalin and Tyutin \cite{bt}, 
Schwarz \cite{schwarz}, and Hata and Zwiebach \cite{hz},
have contributed to a covariant formulation of odd symplectic geometry.

\vs
\noi
The paper is roughly organized as follows: After a short review of 
anti-symplectic geometry, we introduce in 
Section~\ref{almostparitystructure} the notion of an almost parity 
structure. The main idea is that whereas the Grassmann parity in 
general is a property of the local coordinate charts only, we would 
like to ask when it is as a property of the manifold.
The appropriate tool is an almost parity structure.
With an almost parity structure at hand we may introduce various 
new geometric constructions. In Section~\ref{connection} we discuss 
a connection in anti-symplectic geometry.
In Section~\ref{lonekaehlermann} we restrict ourselves to consider 
odd K\"{a}hler manifolds. Finally, we give a vielbein formulation 
in Section~\ref{vielbein}. Super conventions are written down in an 
Appendix.

\subsection{Basic Settings}

\noi
Let the number $2\nnn$ of variables be {\em finite}.
The anti-symplectic phase space $M$ is a $(\nnn | \nnn)$ real superspace
with a non-degenerate antibracket,
\bea
 (F,G) &=&( F \papar{\Gamm^{A}} ) E^{AB} (\papal{\Gamm^{A}}G)~, \cr \cr
 E^{AB} E_{BC}&=&\delta^{A}_{C}~=~E_{CB} E^{BA} ~,
\eea 
or equivalently an anti-symplectic two-form
\beq
E~=~\hf~ d\Gamm^{A}~ E_{AB} \wedge d\Gamm^{B}    
~=~ - \hf~ E_{AB} ~ d\Gamm^{B} \wedge d\Gamm^{A}   ~. 
\eeq 
The Jacobi identity can neatly be restated as that the two-form $E$ 
is closed
\beq
 dE~=~0~.
\eeq
The antibracket has the following symmetry
\beq
( G , F  )~=~ - (-1)^{(\epsilon_{F}+1)( \epsilon_{G}+1)} ( F , G )  ~, 
\eeq
\beq
 \begin{array}{rclcrcl}
 E^{BA}&=&-(-1)^{(\epsilon_A+1)(\epsilon_B+1)} E^{AB} &~,~~& 
\epsilon(E^{AB})&=&\epsilon_A+ \epsilon_B+1~, \cr
 E_{BA}&=& -(-1)^{ \epsilon_A\epsilon_B} 
E_{AB}  &~,~~&   
\epsilon(E_{AB})&=&\epsilon_{A}+ \epsilon_{B}+1 ~.
\end{array}
\label{absym}
\eeq
Locally there exists an anti-symplectic potential 
\mb{\vartheta~=~\vartheta_{A}~d\Gamm^{A}} such that \mb{E=d\vartheta}.
Written out
\beq
 E_{AB}~=~ (\larrow{\partial^{l}_{A}}\vartheta_{B})
+(-1)^{ \epsilon_A\epsilon_B}(\larrow{\partial^{l}_{B}}\vartheta_{A})~.
\eeq
Locally one may resort to anti-symplectic Darboux coordinates
\mb{\Gamm^{A}} \mb{=} \mb{(\phi^{\alpha},\phi^{*}_{\alpha})},
where \mb{\epsilon(\phi^{*}_{\alpha})}
\mb{=}\mb{\epsilon(\phi^{\alpha})+1},
so that
\beq
\begin{array}{rcccl}
E^{\alpha}{}^{*}_{\beta}&=&\delta^{\alpha}_{\beta}
&=&-E_{\beta}^{*}{}^{\alpha} \cr
E^{\alpha\beta}&=&0&=&E^{*}_{\alpha}{}^{*}_{\beta}~,
\end{array}
\eeq
or in terms of the fundamental antibrackets
\beq
(\phi^{\alpha},\phi^{*}_{\beta}) 
=\delta^{\alpha}_{\beta}~,~~~~~~~~~
(\phi^{\alpha},\phi^{\beta}) =0~,~~~~~~~~~
(\phi^{*}_{\alpha},\phi^{*}_{\beta}) =0~.
\eeq

\subsection{Odd Pfaffian}
\label{oddpfff}

\noi
Perhaps one of the most important difference between even versus odd
symplectic geometry is that there is no canonical Liouville 
measure\footnote{In this paper all measure densities \mb{\rho} are 
{\em signed} single-valued densities. This means that the 
supermanifold (and in particular the body) has to be orientable.} 
in odd symplectic geometry. The naive guess would probably be an
odd Pfaffian \mb{{\rm Pf}(E_{AB})} of \mb{E_{AB}}. But it is easy to
see that the volume element cannot be a function of \mb{E_{AB}}.
This is because an anti-symplectic coordinate change 
\mb{\Gamm^{A} \to \Gamm'^{A}(z)},
which by definition leaves \mb{E_{AB}} invariant,
may not be volume preserving, \ie the volume density $\rho$ may changes
although \mb{E_{AB}} is not being changed. 

\vs
\noi
On the other hand, there is a closed one-form
\beq
 {\cal C}~\equiv~ \hf d\Gamm^{A}( \larrow{\partial^{l}_{A}} E_{BC}) E^{CB}
 (-1)^{\epsilon_{B}}~=~0~,
\eeq
which is zero by the Darboux Theorem.
A reasonable definition of the odd Pfaffian \mb{{\rm Pf}(E_{\cdot\cdot})}
should evidently satisfy \mb{{\cal C}=d\ln{\rm Pf}(E_{\cdot\cdot})}. 
So according to this identification 
\mb{{\rm Pf}(E_{\cdot\cdot})} is a {\rm constant}. It therefore 
transforms as a scalar. This is in striking contrast to the even 
symplectic case, where the Pfaffian of the symplectic metric 
transforms as a scalar density.

\setcounter{equation}{0}
\section{Almost Parity Structure}
\label{almostparitystructure}

\subsection{Almost Darboux Coordinates}
\label{almostdarboux}

\noi
Let us introduce the notion of ``almost Darboux coordinates''. These 
are coordinates where all non-vanishing entries of the anti-symplectic 
metric are {\em bosonic}, \ie
\beq
\epsilon(E^{AB})~\equiv~\epsilon_A+\epsilon_B+1~=~0~.
\eeq
A set of Darboux coordinates is an example of almost Darboux coordinates, 
as the name indicates. Let us for completeness mention that under the 
assumption of invertibility, the following conditions are equivalent:
\begin{enumerate}
\item
 $E^{AB}$ is given in almost Darboux coordinates.
\item
 $E^{AB}$ is bosonic.
\item
 $E^{BA}=-E^{AB}$ is antisymmetric.
\item
 $E_{BA}=-E_{AB}$ is antisymmetric.
\item
 $E_{AB}$ is bosonic.
\item
$E^{AB}$ anticommutes with the Grassmann parity operator 
\mb{(-1)^{\epsilon_{A}}}:
\beq
 ((-1)^{\epsilon_{A}}+ (-1)^{\epsilon_{B}}) E^{AB} ~\equiv~
  (-1)^{\epsilon_{A}} E^{AB}+E^{AB}(-1)^{\epsilon_{B}}~=~0~.
\eeq
\end{enumerate} 
\noi
Almost Darboux coordinates are clearly stabile under Grassmann preserving
coordinate changes \mb{\Gamm^{A} \to \Gamm'^{B}(\Gamm)}, \ie when the 
non-vanishing entries of the Jacobian-matrix
\beq 
\Gamm'^{B}\papar{\Gamm^{A}}
\eeq
 has even Grassmann-grading
\beq
\epsilon(\Gamm'^{B}\papar{\Gamm^{A}})~\equiv~\epsilon_{A}+\epsilon_{B}~=~0~.
\label{bostrans}
\eeq

\subsection{Almost Parity Structure}

\noi
The Grassmann parity operator \mb{(-1)^{\epsilon_{A}}} is invariant 
under Grassmann preserving changes of coordinates \eq{bostrans} but not 
under general coordinate transformations. Also let us note that the
Grassmann parity is a property of the coordinates, not the manifold itself.
Let us generalize the Grassmann parity operator \mb{(-1)^{\epsilon_{A}}}
to a covariant object in the following way. Consider a general 
\mb{(\nnn_{+}|\nnn_{-})} supermanifold $M$. An {\em almost parity structure} 
\mb{P:TM \to TM} is a Grassmann-even $(1,1)$-tensor, whose square is the 
identity,
\bea
 P&=& \partial^{r}_{A}~  P^{A}{}_{B}~ \otimes ~ \larrow{d\Gamm^{B}}~,
~~~~~~~~~~\epsilon( P^{A}{}_{B})~=~\epsilon_{A}+\epsilon_{B}~, \cr
\hf[P,P]&=&P^2~=~{\rm Id}
~=~ \partial^{r}_{A}~\otimes~ \larrow{d\Gamm^{A}}~.
\eea
(In the first equality \mb{\partial^{r}_{A}} does {\em not} differentiate 
the \mb{P^{A}{}_{B}}-functions, see the Appendix.)
We will furthermore assume that the supertrace of $P$ is equal to the 
dimension of the manifold:
\beq
 (-1)^{\epsilon_{A}} P^{A}{}_{A}~=~{\rm str} (P) ~=~ \nnn_{+}+\nnn_{-}~.
\label{supertracecond}
\eeq
It is convenient to introduce the idempotent projection operators
\beq
   P_{\pm}~=~ \hf({\rm Id} \pm P)~,~~~{\rm Id}~=~P_{+}+P_{-}~,~~~
P~=~P_{+}-P_{-}~,~~~ P_{\pm}P_{\pm}~=~P_{\pm}~,~~~
P_{\pm}P_{\mp}~=~0~.
\eeq
Consider a point \mb{m \in M} on the manifold, and a coordinate system 
which cover this point $m$. The natural basis of tangent vectors 
in \mb{T_{m}M} wrt.\ the choosen coordinate system is 
\mb{(\partial^{r}_{A})_{A=1,\ldots,\nnn_{+}+\nnn_{-}}}.
However \mb{P_{m}:T_{m}M\to T_{m}M} may not be diagonal in this basis. 
To diagonalize the almost parity structure $P_{m}$, one should perform 
a change of the basis to a new basis 
\mb{(e^{r}_{(A)})_{A=1,\ldots,\nnn_{+}+\nnn_{-}}} for the tangent 
space \mb{T_{m}M}: 
\beq
     \partial^{r}_{A} ~=~ e^{r}_{(B)}~ \Lambda^{B}{}_{A}~,
\eeq
where \mb{\Lambda^{B}{}_{A}} is an invertible matrix. We will restrict 
the allowed basis shift to shift that carries definite Grassmann parity 
\mb{\epsilon(\Lambda^{B}{}_{A})=\epsilon_{A}+\epsilon_{B}}. The almost 
parity structure \mb{P:T_{m}M\to T_{m}M} is diagonalizable in this 
restricted sense with eigenvalues \mb{\sigma=\pm1}. The eigenspaces 
for $P$ are \mb{P_{\pm}(TM)}.  Moreover, we see from the condition 
\eq{supertracecond}, that the eigenspace \mb{P_{+}(TM)} (\mb{P_{-}(TM)}) 
has multiplicity \mb{\nnn_{+}} (\mb{\nnn_{-}}) and their bases carry 
Grassmann parity $0$ ($1$), respectively. As a consequence the 
superdeterminant of $P$ is
\beq
    {\rm sdet} (P) ~=~ (-1)^{\nnn_{-}}~.
\eeq

\vs
\noi
{\bf Proposition.}
{\em Two almost parity structures \mb{P_{(1)}} and \mb{P_{(2)}} are globally
related via similarity transformations, \ie there exists a global 
automorphism \mb{\Lambda:TM \to TM}, such that 
\mb{P_{(1)}\Lambda} \mb{=} \mb{\Lambda P_{(2)}}.}

\vs
\noi
{\em Sketched proof}: To prove the statement locally, we choose a 
coordinate system. It is enough to show that $P$ is related via a 
similarity transformation to the Grassmann parity. This follows from the
discussion above. Finally, the global statement follows by use of a 
partition of the unity.
\proofbox

\vs
\noi
Now consider an odd $(\nnn | \nnn)$ supermanifold.
An almost parity structure $P$ and an anti-symplectic structure $E$
are {\em compatible} with the anti-symplectic structure $E$ iff
\beq
    P^{A}{}_{B}~ E^{BC}~ (P^{T})_{C}{}^{D}~=~-E^{AD}~,
\label{epep}
\eeq
or equivalently
\bea
   (P^{T})_{A}{}^{B} ~ E_{BC}~ P^{C}{}_{D}~=~-E_{AD}~,~~~~~~&&~~~~~
[P\stackrel{\wedge}{,} E]~=~0~,\cr
E(PX,Y)&=&-E(X,PY)~,
\eea
where $X$ and $Y$ are vector fields. Here the super transposed 
\mb{ P^{T} = d\Gamm^{B} (P^{T})_{B}{}^{A} \otimes \larrow{i^{l}_{A}}~
:~T^{*}M \to T^{*}M} is
\beq
(P^{T})_{B}{}^{A}~=~(-1)^{(\epsilon_{A}+1)\epsilon_{B}} P^{A}{}_{B}~.
\eeq
Moreover, \mb{E(P_{\pm}X,Y)~=~E(X,P_{\mp}Y)}.
The eigenspaces \mb{P_{\sigma}(TM)} are Lagrangian subspaces of 
the tangent space \mb{TM}:
\bea
E(P_{\pm}X,P_{\pm}Y)&=&0~, \cr
(P^{T}_{\sigma})_{A}{}^{B}~ E_{BC}~(P_{\sigma})^{C}{}_{D} ~=~0~,~~~~&&~~~~
    (P_{\sigma})^{A}{}_{B}~ E^{BC}~ (P^{T}_{\sigma})_{C}{}^{D}~=~0~.
\label{lagrsubmanifoldcond}
\eea
Note that the compatibility condition \eq{epep} is a non-trivial 
condition even for a cohomology-exact odd (possibly degenerate) two-form. 
Therefore, the condition makes some of the representants in a cohomology 
class \mb{[E]} more preferable.

\vs
\noi
The almost parity structure $P$ is clearly analogous to the almost
complex structure $J$ in complex differential geometry.  
Let us search for the counterparts of the complex conjugation,
the Nijenhuis tensor $N$, the K\"{a}hler potential $K$,
the Dolbeault differentials, etc. A different approach binding together
K\"{a}hler and BV geometry has been studied by Aoyama and Vandoren 
\cite{vandoren} and Khurdaverdian and Nersessian \cite{khunerkaehler}.

\subsection{Canonical Examples}

\vs
\noi
{\em Darboux Coordinates}.
It is perhaps useful to be a bit more explicit in case of an
anti-symplectic \mb{(\nnn | \nnn)} supermanifold with an almost parity
structure $P$. Consider Darboux coordinates. In the usual \mb{2 \times 2} 
Grassmann block representation, we have 
\beq
   P{}^{\cdot}{}_{\cdot}~=~\twobytwo{P_{00}}{P_{01}}{P_{10}}{P_{11}}~,~~~~
   E_{(0)}^{\cdot\cdot}~=~ \twobytwo{0}{{\bf 1}}{-{\bf 1}}{0} ~,~~~~
   P_{(0)}{}^{\cdot}{}_{\cdot}~=~\twobytwo{{\bf 1}}{0}{0}{-{\bf 1}}~,
\eeq
where \mb{P_{(0)}} denote the Grassmann parity.
From the compatible condition \eq{epep} we get that
\beq
    P_{11}=- P_{00}^{T}~,~~~~~~~~
  P_{01}^{T}=-P_{01}~,~~~~~~~~~~P_{10}^{T}=P_{10}~.
\eeq
The supertrace condition \eq{supertracecond} and \mb{P^2={\rm Id}} yields 
that \mb{P_{00}={\bf 1}+S}, where \mb{S} denotes the soul part of 
\mb{P_{00}}. It is straightforward to show by use of \mb{P^2={\rm Id}} that 
\bea
  \twobytwo{{\bf 1}+ S}{P_{01}}{P_{10}}{-{\bf 1}- S^{T}}~ 
 \twobytwo{{\bf 1}+\frac{1}{2} S}{\frac{1}{2}P_{01}}
          {\frac{1}{2} P_{10}}{-{\bf 1}- \frac{1}{2}S^{T}} 
&=&\twobytwo{{\bf 1}+\frac{1}{2} S}{-\frac{1}{2}P_{01}} 
               {\frac{1}{2} P_{10}}{{\bf 1}+ \frac{1}{2}S^{T}}  \cr  \cr
&=& \twobytwo{{\bf 1}+\frac{1}{2} S}{\frac{1}{2}P_{01}} 
               {\frac{1}{2} P_{10}}{-{\bf 1}- \frac{1}{2}S^{T}}~
    \twobytwo{{\bf 1}}{0}{0}{-{\bf 1}} ~.\cr
&&
\eea
So the diagonalizing transformation is \mb{ P\Lambda=\Lambda P_{(0)}} with
\beq
   \Lambda^{\cdot}{}_{\cdot}~=~ 
 \twobytwo{{\bf 1}+\frac{1}{2} S}{\frac{1}{2}P_{01}}
               {\frac{1}{2} P_{10}}{-{\bf 1}- \frac{1}{2}S^{T}}~.
\eeq

\vs
\noi
{\em Almost Darboux Coordinates}.
Let us next consider an example of almost Darboux coordinates 
\mb{\Gamm^{A}=(x^{\alpha};\theta^{\bar{\alpha}})}, where
\mb{\alpha,\bar{\alpha}} \mb{=} \mb{1, \ldots, \nnn}, and where 
\mb{x^{\alpha}} are bosons and \mb{\theta^{\bar{\alpha}}} are fermions. 
The almost parity structure $P$ is assumed to be equal to the Grassmann 
parity \mb{P^{A}{}_{B}=(-1)^{\epsilon_{A}} \delta^{A}_{B}}.
As a consequence the Nijenhuis tensor $N$ for the almost parity structure 
$P$ vanishes (see Section~\ref{nijenhuis}). Because the coordinates are 
almost Darboux, the block diagonal pieces of the anti-symplectic metric 
vanish,
\beq
   E_{\alpha\beta}~=~0~=~E_{\bar{\alpha}\bar{\beta}}~,
\eeq
cf.\ \eq{lagrsubmanifoldcond}. {}From the closeness of the anti-symplectic 
two-form $E$, one may prove that there exists locally a corresponding
odd parity potential \mb{K=K(x,\theta)} such that 
\beq
 - E_{\alpha\bar{\beta}}~=~  E_{\bar{\beta}\alpha}
~=~ (\papal{\theta^{\bar{\beta}} } K \papar{x^{\alpha} }) ~.
\label{ekek}
\eeq
Moreover, any other odd parity potential $K'$ differs from $K$ in the 
following way:
\beq
      K'(x,\theta)-K(x,\theta)~=~F(x) +\bar{F}(\theta)~.
\eeq
The parity potential $K$ is quite similar to the K\"{a}hler potential 
$K$ appearing in standard complex differential geometry. When there is no
confusion, we shall simply call $K$ a K\"{a}hler potential. 
The parity preserving coordinate transformations are transformations 
of the form
\beq
 x^{\alpha} \to x'^{\alpha}(x)~,~~~~~~~~~~~~~~~~~ 
\theta^{\bar{\alpha}} \to \theta'^{\bar{\alpha}}(\theta)~.
\eeq

\vs
\noi
In this example the eigenspaces are:
\beq
 P_{+}(TM)~=~{\rm span}(\papal{x^{\alpha}})~,~~~~~~~~
 P_{-}(TM)~=~{\rm span}(\papal{\theta^{\alpha}})~.
\eeq
The submanifolds 
\beq
\{ (x^{\alpha};\theta^{\bar{\alpha}}) | 
 x^{\alpha}=x_{(0)}^{\alpha} \}~,~~~~~~~
\{(x^{\alpha};\theta^{\bar{\alpha}}) | 
\theta^{\bar{\alpha}}=\theta_{(0)}^{\bar{\alpha}} \}.
\eeq
are two Lagrangian submanifolds of Grassmann parity $(\nnn |0)$ and 
$(0| \nnn)$, respectively, that intersect the point 
\mb{\Gamm_{(0)}^{A}=(x_{(0)}^{\alpha};\theta_{(0)}^{\bar{\alpha}})}.

\vs
\noi
Let us for simplicity restrict to the case of constant almost 
Darboux coordinates 
\beq
E~=~E^{(0)}_{\alpha\bar{\alpha}}~
d\theta^{\bar{\alpha}} \wedge dx^{\alpha}~.
\eeq
The group $G$ of anti-symplectic, parity preserving coordinate 
transformations is a subgroup of the rigid (``global'', ``point 
independent'') transformations. In fact, it is the semidirect product 
of the rigid translations and the rigid $GL(\nnn)$ group of linear 
transformations of the form
\bea
x'^{\alpha}&=&\lambda^{\alpha}{}_{\beta}~ x^{\beta}~, \cr
\theta'^{\bar{\alpha}}&=& E_{(0)}^{\bar{\alpha}\alpha}~ 
(\lambda^{-1,T})_{\alpha}{}^{\beta}~
E^{(0)}_{\beta\bar{\beta}}~ \theta^{\bar{\beta}}~.
\eea
The volume density changes as 
\mb{\rho'={\rm sdet}(\lambda^{\cdot}{}_{\cdot} )^{-2}~ \rho}.
We shall later see that the group $G$ of anti-symplectic, 
parity preserving coordinate transformations may equally well
be characterized as the group of ortho-symplectic transformations.

\section{Parity Structure}

\subsection{Definition of Parity Structure}
\label{parity}

\noi
Consider a \mb{(\nnn_{+}|\nnn_{-})} supermanifold $M$ equipped with an almost 
parity structure $P$. A coordinate system is said to {\em adapt} an almost 
parity structure $P$ if the natural basis of tangent vectors
\mb{\partial_{A}^{r}} are eigenvectors for the almost parity structure $P$ 
with eigenvalue \mb{(-1)^{\epsilon_{A}}}:
\beq
  \partial_{B}^{r} ~P^{B}{}_{A}~\equiv~ P(\partial_{A}^{r})
~=~(-1)^{\epsilon_{A}}\partial_{A}^{r}~.
\eeq
or equivalently
\beq
   (P^T)_{A}{}^{B}~\partial_{B}^{l}~=~(-1)^{\epsilon_{A}}\partial_{A}^{l}~.
\eeq
We shall see in the Section~\ref{nijenhuis}
that the Nijenhuis tensor $N$ corresponding to the almost parity structure 
$P$  vanishes in regions of the manifold that are covered by $P$-adapted 
coordinate patches. We will also see that the two eigenspaces 
\mb{P_{\sigma}(TM)} are stabile under the Lie bracket \mb{[~,~]} in these 
regions.

\vs
\noi
$P$-adapted coordinates are also almost Darboux coordinates,
cf.\ Section~\ref{almostdarboux}, but the opposite need not be the case. 

\vs
\noi
A {\em parity preserving coordinate transformation} is a coordinate 
transformation between two local coordinate system such that the 
non-vanishing entries of the Jacobian-matrix
\beq 
          \Gamm'^{B}\papar{\Gamm^{A}}
\eeq
has positive $P$-parity, \ie is bosonic. So a parity preserving coordinate 
transformation is the same as a Grassmann parity preserving coordinate 
transformation, cf.\ \ref{bostrans}. Note that a coordinate transformation 
\mb{(x^{\alpha}; \theta^{\bar{\alpha}})\to 
(x'^{\alpha}; \theta'^{\bar{\alpha}})} between two $P$-adapted coordinate 
patches is a parity preserving coordinate transformation. Therefore 
\mb{x'^{\alpha}=x'^{\alpha}(x)} and
\mb{\theta'^{\bar{\alpha}}=\theta'^{\bar{\alpha}}(\theta)}. So the only
coordinate transformations among $P$-adapted coordinate charts are a
``holomorphic'' change of the ``holomorphic'' variables \mb{x^{\alpha}} and 
an {\em independent} ``anti-holomorphic'' change of the 
``anti-holomorphic'' variables \mb{\theta^{\bar{\alpha}}}.

\vs
\noi
{\bf Definition}.
An almost parity structure $P$ is a {\em parity structure} iff there 
exists an atlas of $P$-adapted coordinate charts.

\vs
\noi
{\bf Definition}.
An {\em odd almost K\"{a}hler manifold} \mb{M,P,E)} is a 
\mb{(\nnn | \nnn)}-manifold  $M$ with a non-degenerate anti-symplectic 
structure $E$ and an almost parity structure $P$ that is compatible with $E$, 
cf.\ \eq{epep}.  (Actually, this is what one would normally call an odd 
almost {\em pseudo}-K\"{a}hler manifold, but we shall drop the prefix 
{\em pseudo} for convenience.) We will later show that an odd almost 
K\"{a}hler manifold can be equipped with an odd metric (\ie the  odd 
{\em K\"{a}hler} metric) and its Levi-Cevita connection.  

\vs
\noi
{\bf Definition}.
An {\em odd K\"{a}hler manifold} \mb{M,P,E)} is an odd almost 
K\"{a}hler manifold, where the almost parity structure $P$ is a
parity structure. 

\vs
\noi
We conclude that an anti-symplectic $(\nnn | \nnn)$-manifold $M$ possesses 
a compatible Grassmann parity structure iff there exists an atlas of almost 
Darboux charts, which are mutually connected via Grassmann parity 
preserving transformations, see also \eq{bostrans}. 
This means that every point in the manifold is intersected by
a $P$-adapted $(\nnn |0)$ and a $(0| \nnn)$ Lagrangian submanifold,
cf.\ \eq{lagrsubmanifoldcond}.

\vs
\noi
Below follows an analysis of the integrability of an almost parity 
structure $P$.

\subsection{Characteristic one-forms}

\noi
It is convenient to locally introduce a (double) overcomplete generating 
set of vectors \mb{X_{(\sigma,A)}} for the eigenspace \mb{P_{\sigma}(TM)}, 
\mb{\sigma=\pm1}.
\beq
    X_{(\sigma,A)}~=~\partial^{r}_{B}~ (P_{\sigma})^{B}{}_{A}
~=~(-1)^{\epsilon_{A}+\epsilon_{B}}(P^{T}_{\sigma})_{A}{}^{B}~
 \partial^{l}_{B}~,~~~~~\epsilon( X_{(\sigma,A)})~=~\epsilon_{A}~.
\eeq
Analogously we can define the (double) overcomplete generating set of
characteristic one-forms for the eigenspace \mb{P_{\sigma}(TM)}
\beq
    \eta^{(\sigma,A)}~=~(P_{\sigma})^{A}{}_{B}\larrow{d\Gamm^{B}}~,~~~~~
\epsilon( \eta^{(\sigma,A)} )~=~\epsilon_{A}~.
\eeq
Obviously, we have
\beq
   \eta^{(\sigma,A)}( X_{(\tau,B)})
~=~\delta^{\sigma}_{\tau}~ (P_{\sigma})^{A}{}_{B}~,~~~~~~~~
 \eta^{(+,A)}+ \eta^{(-,A)}~=~d\Gamm^{A}~,
\eeq
and
\beq
   \bigcap_{A} {\rm Ker}( \eta^{(\sigma,A)})~=~P_{\sigma}(TM)~.
\eeq
We furthermore define two-forms
\beq
 a^{(A)}~=~ \pm 2 d \eta^{(\pm,A)}
~=~ (P^{A}{}_{B} \rarrow{\partial^{r}_{C}}) ~
d\Gamm^{C} \wedge  d\Gamm^{B}
~=~\hf a^{(A)}{}_{BC} ~d\Gamm^{C} \wedge  d\Gamm^{B}~,
\eeq
where
\beq
   a^{(A)}{}_{BC}~=~(P^{A}{}_{B} \rarrow{\partial^{r}_{C}})
-(-1)^{\epsilon_{B}\epsilon_{C}}(P^{A}{}_{B} \rarrow{\partial^{r}_{C}})~.
\eeq

\subsection{An Odd Parity Conjugation}
\label{oddparityconj}

\noi
Consider an \mb{(\nnn | \nnn)} anti-symplectic supermanifold $M$.
What should be the analogue of complex conjugation? Loosely speaking, it
should be the operation that takes fields \mb{\phi^{\alpha}} into
its corresponding antifield \mb{\phi^{*}_{\alpha}}.
Let us start by defining the concept of odd parity conjugation
in the frame bundle over the manifold. Let 
\mb{(e_{(A)})_{A=1,\ldots, 2\nnn}} be a basis for the tangent space $TM$.
Then the parity conjugated basis \mb{(e_{*}^{(A)})_{A=1,\ldots, 2\nnn}}
(of opposite Grassmann parity) is the unique basis such that
\beq
E(e_{(A)},e_{*}^{(B)})~=~\delta^{B}_{A}~.
\eeq
Applying conjugation twice yields a minus
\beq
e_{(A)}^{**}~=~-e_{(A)}~.
\eeq
For a tangent vector \mb{X=X^{A} e_{(A)}} we define a parity 
conjugated tangent vector \mb{X^{*}=X^{A} e_{*}^{(A)}}. 
This of course depends strongly on the choice of basis.
Also the notion of real and imaginary part has no counterpart.
The only nice things to say is that no matter which basis we choose,
the conjugation is linear, \mb{X^{**}=-X}.
If $E$ and $P$ are compatible, and \mb{(e_{(A)})_{A=1,\ldots, 2\nnn}}
are eigenvectors for $P$, then \mb{(e_{*}^{(A)})_{A=1,\ldots, 2\nnn}}
are also eigenvectors for $P$ (with opposite eigenvalue), so that
in this case the $*$-conjugation gives a bijection 
between \mb{P_{+}(TM)} and \mb{P_{-}(TM)}. We emphasize that 
\mb{[X,Y]^{*}=[X^{*},Y^{*}]} does {\em not} hold in general.

\vs
\noi
For an  one-form \mb{\eta} the parity conjugated one-form \mb{\eta^{*}} is 
defined via \mb{\eta^{*}(X)~=~\eta(X^{*})}.
The above definitions leads to the convenient rules
\beq
\left(\papal{\phi^{\alpha}} \right)^{*}
~=~\papal{\phi^{*}_{\alpha}}~,~~~~~~~~~~
(d\phi^{\alpha})^{*}~=~\phi^{*}_{\alpha}~,
\eeq
in Darboux coordinates \mb{E~=~d\phi^{*}_{\alpha}\wedge d\phi^{\alpha}}.

\subsection{Nijenhuis Tensor}
\label{nijenhuis}

\noi
We start by defining two tensors \mb{N_{\sigma}~:~TM \times TM \to TM},
\beq
 N_{\pm}(X,Y)~=~P_{\mp}[P_{\pm}X,P_{\pm}Y]
 ~=~-(-1)^{\epsilon(X)\epsilon(Y)}N_{\pm}(Y,X)~.
\eeq 
Note that \mb{ N_{\pm}(X,PY)~=~\pm N_{\pm}(X,Y)~=~N_{\pm}(PX,Y)}.
One can write the Lie-bracket symbol \mb{[~,~]~:~TM \times TM \to TM}
in the following way
\beq
[~,~]~=~~ \larrow{d\Gamm^{A}}~ \times ~\papal{\Gamm^{A}}
-(-1)^{\epsilon_{A}}\papal{\Gamm^{A}}~ \times ~\larrow{d\Gamm^{A}}~.
\eeq
Then we can write 
\mb{N_{\pm}~\in~\Gamma(TM \otimes \Lambda^{2}(T^{*}M))} as
\beq
N_{\pm}~=~P_{\mp}\left( [~,~](P_{\pm} \times P_{\pm}) \right)
~=~ \hf \partial^{r}_{A}~N_{\pm}^{A}{}_{BC}~
\larrow{d\Gamm^{C}} \wedge  \larrow{d\Gamm^{B}}~.
\eeq
We now define the Nijenhuis tensor as
\beq
 N~=~4(N_{+}+N_{-})~,
\eeq
or
\bea
N(X,Y)&=&[X,Y]+[PX,PY]-P[X,PY]-P[PX,Y] \cr
&=&-(-1)^{\epsilon(X)\epsilon(Y)}N(Y,X)~.
\eea
It satisfies \mb{N(PX,Y)~=~N(X,PY)}.
In components the Nijenhuis tensor 
\beq
N~=~\hf \partial^{r}_{A}~N^{A}{}_{BC}~
\larrow{d\Gamm^{C}} \wedge  \larrow{d\Gamm^{B}}~
\in~\Gamma(TM \otimes \Lambda^{2}(T^{*}M))
\eeq
reads
\bea
 \larrow{d\Gamm^{A}}(N(\partial^{r}_{B},\partial^{r}_{C}))
&=& - N^{A}{}_{BC}   \cr
&=& \left(P^{A}{}_{D}~(P^{D}{}_{B}\rarrow{\partial^{r}_{C}})
- (P^{A}{}_{B}\rarrow{\partial^{r}_{D}})~P^{D}{}_{C} \right)
-(-1)^{\epsilon_{B}\epsilon_{C}}(B \leftrightarrow C)~.
\eea 
The relation can be inverted to give
\bea
 8 N_{\pm}(X,Y)&=&N(X,Y) \pm N(X,PY)~=~2 N(X,P_{\sigma}Y) ~, \cr
8 N^{A}_{\sigma}{}_{BC}
&=&N^{A}{}_{DB}~(P_{\sigma})^{D}{}_{C}
-(-1)^{\epsilon_{B}\epsilon_{C}}(B \leftrightarrow C)  ~.
\eea
We observe that \mb{(-1)^{\epsilon_{A}}N^{A}{}_{AB}=0} and 
\mb{(-1)^{\epsilon_{A}}P^{A}{}_{B}N^{B}{}_{AC}=0}.

\subsection{Integrability Condition}

\noi
By the Frobenius Theorem, an almost parity structure is locally a parity 
structure if one of the following equivalent integrability criteria 
is satisfied: 
{\em \begin{enumerate}
\item
The Nijenhuis tensor \mb{N=0} vanishes.
\item
Both of the tensors \mb{N_{\pm}=0} vanishes.
\item
Both of the eigenspaces \mb{P_{\sigma}(TM)} is stabile 
under the Lie-bracket operation \mb{[~,~]}.
\item
\mb{\forall \sigma}~:~The ideal \mb{{\cal I}(\eta^{(\sigma,A)} )} in the 
exterior algebra of forms, generated by the characteristic one-forms 
for \mb{P_{\sigma}(TM)}, is stabile under the action of the exterior 
derivative $d$.
\end{enumerate}} 

\noi
We conjecture that a vanishing Nijenhuis tensor actually ensures
that the manifold admits the parity structure {\em globally}.
The corresponding statement for bosonic complex manifolds
was proven by Newlander and Nirenberg \cite{newnir}.

\subsection{Dolbeault Bi-Complex}
\label{dollydagger}

\noi
In this Subsection we review some basic facts about Dolbeault bi-complexes,
that we need later.
Consider some $P$-adapted coordinates \mb{(x^{\alpha};\theta^{\bar{\alpha}})}
on a supermanifold $M$ with a parity structure $P$.
We can then split the exterior derivative \mb{d=\partial+\bar{\partial}} 
into its Dolbeault components
\beq
 \partial~=~dx^{\alpha} \papal{x^{\alpha}}~,~~~~~~~~~~~
 \bar{\partial}~=~d\theta^{\bar{\alpha}} \papal{\theta^{\bar{\alpha}}}~.
\eeq
Now let us mention some basic applications of a bigraded complex.

\vs
\noi
{\bf Proposition}.
{\em Consider a zero-form $F$, satisfying 
\mb{\bar{\partial}\wedge \partial F=0}. 
Then \mb{F(x,\theta)=f(x)+\bar{f}(\theta)} splits locally into a 
``holomorphic'' and an ``anti-holomorphic'' piece. This split can be made 
global if the first \v{C}ech cohomology class 
\mb{\check{H}^{1}_{\rm const} (M,\R)=0} vanishes. }

\vs
\noi
(The function \mb{\bar{f}} has no relation to $f$.) The space 
\mb{\check{H}^{1}_{\rm const} (M,\R)} refers to the \v{C}ech cohomology in 
terms of (locally) {\em constant} sections. Moreover, we will work in a 
{\em fixed} atlas of open neighbourhoods \mb{(U_{i})_{i \in I}} covering 
$M$. It is clear that a manifest geometric treatment has to be independent 
of the atlas. But it turns out that this is merely a technical complication 
that we will not discuss here.

\vs
\noi
{\em Proof}: Locally two intersecting neighbourhoods \mb{U_{i}} 
and \mb{U_{j}} defines a one-cochain 
\mb{c^{(1)}~=~\frac{1}{2}c_{ij}di \wedge dj} by a 
``separation of variables'' argument 
\beq
   f_{i}(x)+\bar{f}_{i}(\theta)~=~F(x,\theta)
~=~f_{j}(x)+\bar{f}_{j}(\theta)~
\eeq
so that
\beq
  f_{i}(x)-f_{j}(x)~=~c_{ij}
~=~-\bar{f}_{i}(\theta)+\bar{f}_{j}(\theta)~.
\eeq
The \mb{c^{(1)}} is a one-cochain both in terms of sheaf of general 
sections and in the sense of sheaf of constant sections. We will mainly 
work in the latter. However in the former, \mb{c^{(1)}} is also a 
one-coboundary, so that it is trivially a one-cocycle. Therefore, it is
a (locally) constant one-cycle. By the assumptions it is then a 
one-coboundary in the constant sense, \ie there exists a constant 
zero-cochain \mb{c^{(0)}=c_{i}di} such that \mb{c_{ij}=c_{i}-c_{j}}. 
We may then define \mb{f(x)=f_{i}(x)-c_{i}}.
\proofbox

\vs
\noi
{\bf Proposition}. {\em Consider an exact \mb{(1,1)} two-form 
\mb{\omega^{(11)}=d \eta}, where \mb{\eta=\eta^{(10)}+\eta^{(01)}} is a 
global one-form. If the \v{C}ech cohomology classes
\mb{\check{H}^{1}(M,\R)=0} and
\mb{\check{H}^{2}_{\rm const}(M,\R)=0} vanish, then there exists a global 
zero-form $F$ such that \mb{\omega^{(11)}=(\bar{\partial}\wedge\partial F)}.}

\vs
\noi
{\em Proof}: 
Locally, there exist zero-forms \mb{h_{i}} and \mb{\bar{h}_{i}} such that
\mb{\eta^{(10)}=\partial h_{i}} and 
\mb{\eta^{(01)}=\bar{\partial} \bar{h}_{i}}.
So locally the zero-form \mb{F_{i}=h_{i}-\bar{h}_{i}} will be a solution
to \mb{d\eta=\bar{\partial}\wedge \partial F_{i}}.
Furthermore, two intersecting neighbourhoods \mb{U_{i}} 
and \mb{U_{j}} defines a one-coboundary 
\mb{F^{(1)}~=~\frac{1}{2}F_{ij}di \wedge dj} by
\mb{F_{ij}=F_{i}-F_{j}}. Note that the functions \mb{F_{ij}} satisfy
\mb{\bar{\partial}\wedge\partial F_{ij}=0}. So locally they split 
\mb{F_{ij}(x,\theta)=f_{ij}(x)+\bar{f}_{ij}(\theta)} into 
``holomorphic'' and ``anti-holomorphic'' parts. By maybe going to a
finer atlas we may assume that the splitting is defined on the whole 
\mb{U_{i} \cap U_{j}}. Therefore,  we have one-cochains
\mb{f^{(1)}~=~\frac{1}{2}f_{ij} di \wedge dj} and
\mb{\bar{f}^{(1)}~=~\frac{1}{2}\bar{f}_{ij} di \wedge dj}.
Unfortunately, they are not one-cocycles. But this turns out to
be easy to fix. Let us define a two-cochain
\beq
    c^{(2)}~=~\frac{1}{3!} c_{ijk}di \wedge dj \wedge dk
~=~df^{(1)}~=~-d\bar{f}^{(1)}~,
\eeq
that is \mb{c_{ijk}=f_{ij}(x)+f_{jk}(x)+f_{ki}(x)}. 
Note that \mb{c^{(2)}} is constant by a 
``separation of variable'' argument.
\mb{c^{(2)}} is actually a constant two-cocycle, and therefore
by assumption a two-coboundary in the constant sense,
\ie there exists a  constant one-cochain 
\mb{c^{(1)}=\frac{1}{2} c_{ij }di \wedge dj} such that
\mb{c^{(2)}=dc^{(1)}} or written out \mb{c_{ijk}=c_{ij}+c_{jk}+c_{ki}}.
So we may construct ``holomorphic'' and ``anti-holomorphic''
one-cocycles \mb{f'^{(1)}=f^{(1)}-c^{(1)}} and 
\mb{\bar{f}'^{(1)}=\bar{f}^{(1)}+c^{(1)}}. By assumption there exist
zero-cochains \mb{f^{(0)}} \mb{=} \mb{f_{i}(x,\theta) di } and 
\mb{\bar{f}^{(0)}} \mb{=} \mb{\bar{f}_{i}(x,\theta) di }, such that 
\mb{f^{(1)}=d f^{(0)}} and \mb{\bar{f}^{(1)}=d \bar{f}^{(0)}}.
Now let \mb{F=F_{i}-f_{i}-\bar{f}_{i}}.
\proofbox

\setcounter{equation}{0}
\section{Connection}
\label{connection}

\subsection{Odd Metric}

\noi {\em Unless otherwise stated, we consider in this 
Section~\ref{connection} a $(\nnn|\nnn)$ supermanifold $M$ with a
non-degenerate anti-symplectic structure $E$ and an almost parity 
structure $P$, that are mutually compatible, cf.\ \eq{epep}}
Let us introduce a non-degenerate odd metric $(0,2)$-tensor
\beq
 g ~=~ \hf d \Gamm^{A} ~ g_{AB} \vee d \Gamm^{B}
 ~=~ \hf  g_{AB}~ d \Gamm^{B} \vee d \Gamm^{A}~,
\eeq
by
\bea
   g_{AB}&=&E_{AC}~ P^{C}{}_{B}
~=~(-1)^{\epsilon_{A}\epsilon_{B}} g_{BA}~,~~~~
\epsilon( g_{AB})~=~\epsilon_{A}+ \epsilon_{B}+1~, \cr
&&~~~~~~~~~g~=~ \hf [E\stackrel{\vee}{,} P]~.
\eea
The  symbol $\vee$ is the super-symmetrized tensor product,
\mb{ d \Gamm^{B} \vee d \Gamm^{A}
 = (-1)^{\epsilon_{A}\epsilon_{B}} d \Gamm^{A} \vee d \Gamm^{B}}.
The above symmetry property of $g$ follows from \eq{epep}.
(As we are not demanding any positivity properties of $g$, we should 
properly refer to $g$ as an odd {\em pseudo}-Riemannian metric.)
The inverse metric satisfies
\beq
  g^{BA}~=~(-1)^{(\epsilon_{A}+1)(\epsilon_{B}+1)} g^{AB}~,~~~~~ 
\epsilon(g^{AB})~=~\epsilon_{A}+ \epsilon_{B}+1~. 
\eeq
On the other hand we get
\beq
   (P^{T})_{A}{}^{B} ~ g_{BC}~ P^{C}{}_{D}~=~-g_{AD}~,~~~~~~~~~~~
[P\stackrel{\vee}{,} g]~=~0~.
\eeq
and
\beq
    (P_{\sigma})^{A}{}_{B}~ g^{BC}~ (P^{T}_{\sigma})_{C}{}^{D}~=~0~.
\eeq
Let us note for later that 
\mb{g(\partial^{r}_{A},\partial^{r}_{B})=g_{AB}} and 
\mb{(-1)^{\epsilon_{B}}N^{A}{}_{BC}g^{CB}=0}.
The metric $g$ leads to a $g$-bracket (which we also denote by $g$):
\beq
  g(F,G)~=~ ( F \papar{\Gamm^{A}} ) g^{AB} (\papal{\Gamm^{A}}G)~,
\eeq
where \mb{F,G \in C^{\infty}(M)} are functions.

\subsection{Connection}
\label{conparity}

\noi
We define a Grassmann-even linear connection 
\mb{\nabla:TM \times TM \to TM} in the tangent bundle by
\beq
  \nabla_{X}~=~X^{A} ~\nabla^{l}_{A}~,~~~~~~~~~~~~~~~
  \nabla^{l}_{A}~=~ \papal{\Gamm^{A}}~
+ ~\partial^{r}_{B}~ \Gamma^{B}{}_{AC}~ \larrow{d \Gamm^{C}}~,
\eeq
or equivalently 
\mb{\nabla=d+\Gamma=d\Gamm^{A}\otimes \nabla^{l}_{A}~
:~\Gamma(TM) \to \Gamma(T^{*}M \otimes TM)} where
\beq
\Gamma~=~d\Gamm^{A}~\otimes~\partial^{r}_{B}~\Gamma^{B}{}_{AC}
~\larrow{d\Gamm^{C}}~.
\eeq
Indices $B$ and $C$ are bundle indices, while index $A$ is a 
base manifold index. This distinction is in general useful but 
becomes somewhat blurred for connections over the tangent bundle.  
Acting on forms it reads
\beq
\nabla^{l}_{A} ~=~ \papal{\Gamm^{A}}~
+~:\larrow{i^{r}_{B}}~ \Gamma^{B}{}_{AC}~ d\Gamm^{C}:
~=~ \papal{\Gamm^{A}}~
-(-1)^{\epsilon_{A}\epsilon_{B}} 
\Gamma^{B}{}_{AC}~ d\Gamm^{C} ~\larrow{i^{l}_{B}}~.
\eeq
The existence of two different connections \mb{\nabla^{(1)}} and
\mb{\nabla^{(2)}} reflects the existence of a global non-trivial 
\mb{(1,2)} tensor. The torsion tensor \mb{T:TM \times TM \to TM}
of a connection \mb{\nabla}
\beq
 T(X,Y)~=~\nabla_{X}Y-(-1)^{\epsilon(X)\epsilon(Y)}\nabla_{Y}X-[X,Y]
~=~-(-1)^{\epsilon(X)\epsilon(Y)}T(Y,X)
\eeq
can be viewed as an element 
\beq
T~=~[\nabla\stackrel{\wedge}{,} {\rm Id} ]
~=~[\nabla \stackrel{\wedge}{,} \partial^{r}_{A} \otimes  d\Gamm^{A}]
~\in~\Gamma(TM \otimes \Lambda^{2}(T^{*}M))~.
\eeq
In terms of the Christoffel symbols this means
\beq
 \Gamma^{C}{}_{BA}
~=~-(-1)^{(\epsilon_{A}+1)(\epsilon_{B}+1)}\Gamma^{C}{}_{AB}~.
\label{reallynotorsion}
\eeq
Perhaps the various relations between a connection and an 
almost parity structure are best summarized as
\beq
  \nabla P~=~0~~~~~\Rightarrow~~~~~ P_{\mp} \nabla_{P_{\pm}X}{P_{\pm}Y}~=~0~,
\eeq
\beq
\begin{array}{rclrclrcl}
\cr \cr P\nabla_{X} Y  &=&\nabla_{X} PY && \Leftrightarrow 
&& P_{\pm}\nabla_{X} P_{\mp}Y  &=&0  \cr \cr
 &\Updownarrow &\cr \cr
\nabla P &=&0\cr \cr
 &\Downarrow &\cr \cr
[\nabla\stackrel{\wedge}{,} P ]&=&0\cr \cr 
    &\Updownarrow & ({\rm if}~ T~=~0) \cr \cr 
\nabla_{PX}Y &=& \nabla_{X} PY      && \Leftrightarrow   && 
\nabla_{P_{\pm}X} P_{\mp}Y &=& 0  \cr \cr 
  &&&&&&  &\Downarrow &({\rm if}~ T~=~0)  \cr \cr 
[PX,Y]&=&[X,PY] && \Leftrightarrow   && 
 [P_{\pm}X,P_{\mp}Y]&=&0~, 
\end{array}
\eeq
and
\beq
\begin{array}{c}
N~=~0\cr \cr 
\Uparrow \cr \cr 
\nabla P~=~0~~~~~\wedge~~~~~T~=~0 \cr \cr 
    \Updownarrow \cr \cr 
\nabla P~=~0~~~~~\wedge~~~~~ \nabla_{PX} Y~=~\nabla_{X}PY
 ~~~~~\wedge~~~~~  T~=~0 \cr\cr 
    \Downarrow \cr \cr 
\nabla P~=~0~~~~~\wedge~~~~~ \nabla_{PX} Y~=~\nabla_{X}PY  \cr\cr 
    \Updownarrow \cr\cr 
P_{\rho} \nabla_{P_{\sigma}X}P_{\tau}Y~=~0~~ {\rm vanishes~ for~ mixed~
signs}~~ \rho,~ \sigma,~\tau \cr \cr 
    \Downarrow \cr \cr 
{\rm Holonomy~Group}~~G \subseteq GL(\nnn)~.
\end{array}
\eeq
In a $P$-adapted coordinate system the next-to-last 
condition simply state that the only non-vanishing components of the 
Christoffel symbols \mb{\Gamma^{B}{}_{AC}} are the bose-bose-bose and 
the fermi-fermi-fermi components.

\subsection{Symplectic Connection}
\label{symplcon}

\vs
\noi
A connection is called {\em symplectic} iff it preserves the
anti-symplectic structure:
\beq
 \nabla E ~=~0~,
\label{reallycove}
\eeq
or in components
\beq
  (-1)^{\epsilon_{A}}(\larrow{\partial^{l}_{A}}E_{BC})
~=~(-1)^{\epsilon_{A}\epsilon_{B}} E_{BD} \Gamma^{D}{}_{AC}
-(-1)^{\epsilon_{C}(\epsilon_{A}+\epsilon_{B})} E_{CD} \Gamma^{D}{}_{AB}~.
\eeq
\beq
 (\larrow{\partial^{l}_{A}}E^{BC})
+(-1)^{\epsilon_{A}\epsilon_{B}} \Gamma^{B}{}_{AD} E^{DC}
+(-1)^{\epsilon_{C}(\epsilon_{A}+\epsilon_{B}+1)+\epsilon_{B}} 
\Gamma^{C}{}_{AD} E^{DB}~=~0~.
\label{cove}
\eeq
This condition does not determine uniquely the connection,
not even if we impose that the connection should be torsion-free.
It clearly implies the weaker super-symmetrized condition:
\beq
 [\nabla\stackrel{\wedge}{,} E] ~=~0~~.
\eeq
Applying the Jacobi identity (\mb{dE=0}), this becomes
\beq
  \sum_{{\rm cycl.}~A,B,C} (-1)^{\epsilon_{A}\epsilon_{C}} E_{AD}
\left((-1)^{\epsilon_{B}}  \Gamma^{D}{}_{BC}
-(-1)^{(\epsilon_{B}+1)\epsilon_{C}} \Gamma^{D}{}_{CB} \right)~=~0~.
\eeq
Note that the last equation is satisfied for a torsion-free connection.
We also have that
\beq 
 \nabla P~=~0~~\Leftrightarrow~~\nabla g ~=~0~~\Rightarrow~~
 [\nabla\stackrel{\vee}{,} g] ~=~0 ~.
\eeq

\subsection{Levi-Civita Connection}

\noi  
We are led to consider the Levi-Civita connection. It is the unique 
connection that satisfies
\beq
  T~=~0 ~,~~~~~~~~~~~~~~~~~~~~\nabla g~=~0~, 
\label{notorsion}
\eeq
where $g$ in principle could be any odd, non-degenerate, symmetric metric.
In terms of the Christoffel symbols the metric condition reads
\beq
  (-1)^{\epsilon_{A}}(\larrow{\partial^{l}_{A}}g_{BC})
~=~(-1)^{\epsilon_{C}(\epsilon_{A}+\epsilon_{B})} g_{CD} \Gamma^{D}{}_{AB}
+(-1)^{\epsilon_{A}\epsilon_{B}} g_{BD} \Gamma^{D}{}_{AC}~.
\eeq
This can easily be inverted to yield the familiar formula
\bea
 \Gamma^{D}{}_{AB}&=& (-1)^{\epsilon_{A}}g^{DC}~ \Gamma_{C,AB} ~,~~~~~~~~~
\Gamma_{C,AB}~=~ (-1)^{\epsilon_{A}\epsilon_{B}}\Gamma_{C,BA} ~,\cr
   2\Gamma_{C,AB}
&=&(-1)^{\epsilon_{C}\epsilon_{A}} (\larrow{\partial^{l}_{A}}g_{CB})
+(-1)^{\epsilon_{B}(\epsilon_{A}+\epsilon_{C})}
 (\larrow{\partial^{l}_{B}}g_{CA})
-(\larrow{\partial^{l}_{C}}g_{AB})~.
\label{famlevicivita}
\eea
In a $P$-adapted coordinate system the parity-mixed components 
of the Christoffel symbols \mb{\Gamma^{D}{}_{AB}} vanish.
The following formulas apply to the non-vanishing components only:
\bea
\Gamma^{D}{}_{AB} &=&(-1)^{\epsilon_{A}}g^{DC}
(g_{CA} \rarrow{\partial^{r}_{B}}) 
~=~(-1)^{\epsilon_{A}(\epsilon_{C}+1)}g^{DC} 
(\larrow{\partial^{l}_{A}}g_{CB})~,\cr
g_{AB}&=&(\larrow{\partial^{l}_{A}}K\rarrow{\partial^{r}_{B}})~,\cr
\Gamma_{C,AB}&=&(K\rarrow{\partial^{r}_{C}} \rarrow{\partial^{r}_{A}}
\rarrow{\partial^{r}_{B}})~,   
\eea
where $K$ is a local parity potential.
A connection in the field-sector of the field-antifield space 
has previously been discussed by Alfaro and Damgaard \cite{aldamg} 
in the context of BV quantization of quantum field theories
using Schwinger-Dyson equations.

\subsection{Divergence}

\noi
We define the divergence \mb{{\rm div}(X)} of a bosonic vector field $X$ as
\beq
  {\rm div}(X)~=~{\rm str}(\nabla(X))
~=~ (-1)^{\epsilon_{A}}(\papal{\Gamm^{A}} + F_{A}) X^{A}~,
\eeq
where the contracted Christoffel symbols $F_{A}$ for a general connection 
on the tangent bundle read
\beq
 (-1)^{\epsilon_{A}} F_{A}~=~ \Gamma^{B}{}_{BA}~.
\eeq
$F_{A}$ is not a tensor. From the gauge transformation like property of the 
Christoffel symbols it follows that under a coordinate transformation 
\mb{\Gamm^{A} \to \Gamm'^{A}(\Gamm)} the contracted Christoffel symbols
transforms as 
\beq
   F_{A}~=~(\papal{\Gamm^{A}} \Gamm'^{B}) F'_{B}
 + (\papal{\Gamm^{A}} \ln {\rm sdet}
(\papal{\Gamm^{\cdot}} \Gamm'^{\cdot} ) )    ~.
\eeq 
So \mb{\nabla^{F}=d+F} with \mb{F=d\Gamm^{A}F_{A}} is an (Abelian) 
connection in the superdeterminant bundle over the manifold.
\mb{d-F} is a connection in the inverse superdeterminant bundle,
or equivalently the bundle of volume densities.
In general, the presence of two different superdeterminant connection
fields \mb{F^{(1)}} and \mb{F^{(2)}} witness the existence of a non-trivial
global Grassmann-even one-form \mb{\eta=F^{(2)}-F^{(1)}}.
Notice that global Grassmann-even one-forms \mb{\eta} by the
symplectic metric is in one-to-one correspondance with global Grassmann-odd
vector-fields \mb{V}. We may form another superdeterminant connection
\mb{\nabla^{\FF}} \mb{=} \mb{d+\FF} \mb{=} \mb{d+d\Gamm^{A}\FF_{A}}
with connection field
\beq
    \FF_{A}~=~ (-1)^{(\epsilon_{A}+1)\epsilon_{B}} \Gamma^{B}{}_{AB}~.
\eeq
So \mb{\FF-F} is a globally defined one-form. In the torsion-free case, 
this one-form vanishes identically \mb{F_{A}=\FF_{A}}.

\vs
\noi
In the case of a Levi-Civita connection, we can rewrite the 
contracted Christoffel symbols $F_{A}$ as
\beq
 (-1)^{\epsilon_{A}} F_{A}
~=~ - (-1)^{\epsilon_{B}} g^{BC}  (\larrow{\partial^{l}_{C}}g_{BA}) 
~=~ - (-1)^{\epsilon_{C}}  (\larrow{\partial^{l}_{C}}g^{CB}) g_{BA} ~.
\eeq
Therefore the divergence takes the form
\beq
  {\rm div}(X)
~=~ -(-1)^{\epsilon_{A}}g^{AB}(\papal{\Gamm^{B}}(g_{AC}X^{C}))~.
\eeq
Thus the Levi-Civita divergence 
\beq
{\rm div}(Y_{F})~=~0 
\label{notinterest}
\eeq 
vanishes identically for a vector field of the form 
\beq
Y_{F}~=~g(F,\cdot)~=~g(\cdot,F)~,
\eeq
where $F$ is an odd function. 
However, contrary to even Riemannian geometry where a canonical 
volume density is \mb{{\rm Pf}(g_{AB})}, a volume density in
odd Riemannian geometry is not a function of the metric \mb{g_{AB}}.
There are Killing symmetries which are not volume preserving. 
See also the analogous discussion in Section~\ref{oddpfff}.

\subsection{Odd Laplacian}

\noi
The odd Laplacian $\Delta(F)$ of an odd function $F$ is \cite{khuner}
\beq
  \Delta(F)~=~ - \hf {\rm div}(X_{F})~,
\eeq	
where \mb{X_{F}~=~(F, \cdot)~=~-( \cdot,F)} denotes the Hamiltonian vector 
field for $F$.
We have the following commuting diagram
\beq
\begin{array}{rcccl}
C^{\infty}_{\rm odd}(M) &\stackrel{X_{(.)}}{\longrightarrow}&
\Gamma_{\rm even}(TM) &\stackrel{\nabla}{\longrightarrow}&
\Gamma_{\rm even}(T^{*}M \otimes TM) \cr \cr 
&\searrow& {\rm div} \downarrow & \swarrow& \cr
& -2\Delta~~~~~~&&~~~~~~ {\rm str} & \cr
&& C^{\infty}_{\rm even}(M) &&\end{array}
\label{comdiag}
\eeq
In components $\Delta$ reads \cite{bt,schwarz}
\beq
\Delta~=~\hf (-1)^{\epsilon_{A}}
(\larrow{\partial^{l}_{A}}+F_{A}) E^{AB} \larrow{\partial^{l}_{B}}
~=~\hf (-1)^{\epsilon_{A}} E^{AB}
(F^{(0)}_{B}+\larrow{\partial^{l}_{B}})\larrow{\partial^{l}_{A}}~,
\label{sonice}
\eeq
where \mb{F^{(0)}_{A}} is defined as
\beq
F^{(0)}_{A}~=~-E_{AB}~\Gamma^{B}{}_{CD}~E^{DC}~.
\label{alsonice}
\eeq
The last equality in \eq{sonice} does in general only hold for a
symplectic connection. To derive it, we used the following contracted 
version of \eq{cove}:
\beq
(-1)^{\epsilon_{A}} ((\larrow{\partial^{l}_{A}}+F_{A})E^{AB})
~=~(-1)^{\epsilon_{A}}F^{(0)}_{A}~E^{AB}
~=~(-1)^{\epsilon_{B}}E^{BA}~F^{(0)}_{A} ~.
\label{contractedcove}
\eeq
Let us mention that \mb{F^{(0)}_{A}} has non-trivial transformation 
properties. However, it vanishes identically in $P$-adapted coordinates.
This fact is essential for odd K\"{a}hler manifolds (see 
Section~\ref{lonekaehlermann}).  

\subsection{Analysis of Odd Laplacian}

\noi
The odd Laplacian $\Delta$ is a second order operator, 
or equivalently
\beq
 [[[\stackrel{\rightarrow}{\Delta}, A_{1}], A_{2}], A_{3}]~=~0~,
\eeq
where \mb{A_{1}}, \mb{A_{2}} and \mb{A_{3}} are functions (\ie operators 
of order $0$). The supercommutator 
\mb{\Delta^2} \mb{=} \mb{\hf[\Delta,\Delta]} 
of the second order operator $\Delta$ is at most of order \mb{2+2-1=3}.
In fact, the Jacobi identity guarantees that it is at most of order $2$. 
We can give a proof which does not use the explicite form of $\Delta$,
but merely that it is of second order and Grassmann-odd.
First, note that \cite{beringalg}
\beq
[[[[\Delta,\Delta],A_{1}],A_{2}],A_{3}]~=~
  \sum_{\pi \in {\cal S}_3 } (-1)^{\epsilon_{\pi}}
[[\Delta,[[\Delta,A_{\pi(1)}],A_{\pi(2)}]],A_{\pi(3)}]~=~0~.
\label{comcom}
\eeq 
Here \mb{\epsilon_{\pi}} is a Grassmann factor arising when permuting
\beq
A_{1}~ A_{2}~ A_{3}
~=~(-1)^{\epsilon_{\pi}} A_{\pi(1)}~ A_{\pi(2)}~ A_{\pi(3)}~.
\eeq
Recalling that the antibracket can be expressed as a
multiple commutator \cite{beringalg}
\beq
 (A,B)~=~ (-1)^{\epsilon(A)} [[\stackrel{\rightarrow}{\Delta},A],B]1~,
\eeq
we recognize the right hand side of \eq{comcom} as the Jacobi identity.

\subsection{Curvature}
\label{curvaturesec}

\noi
The {\em curvature tensor} $R$ reads
\bea
 R~=~\hf [\nabla \stackrel{\wedge}{,}\nabla]
&=&-\hf d\Gamm^{D} \wedge d\Gamm^{A}~\otimes~
[\nabla_{A},\nabla_{D}]\cr
&=&-\hf d\Gamm^{D} \wedge d\Gamm^{A}~\otimes~
\partial^{r}_{B}~R^{B}{}_{ADF}~d\Gamm^{F} \cr
&=&- d\Gamm^{D} \wedge d\Gamm^{A}~\otimes~\partial^{r}_{B} \left( 
(-1)^{\epsilon_{A}\epsilon_{B}}(\larrow{\partial^{l}_{A}}\Gamma^{B}{}_{DF})
+ \Gamma^{B}{}_{AC}\Gamma^{C}{}_{DF} \right)\otimes d\Gamm^{F}~, \cr
R^{B}{}_{ADF} &=&(-1)^{\epsilon_{A}\epsilon_{B}}
(\larrow{\partial^{l}_{A}} \Gamma^{B}{}_{DF})
+ \Gamma^{B}{}_{AC}\Gamma^{C}{}_{DF} 
-(-1)^{\epsilon_{A}\epsilon_{D}}(A\leftrightarrow D)~,
\eea
where it is implicitly  understood that there is no contractions
among the base manifold indices, in this case index $A$ and index $D$.
An alternative way of saying this is that one should project
to the appropriated space of base manifold two-forms.
The {\em Ricci tensor} is usually defined as a contraction of a
bundle and a base manifold index with the metric
\beq
 {\rm Ric}~=~d\Gamm^{A}\otimes~\partial^{r}_{B}~{\rm Ric}^{B}{}_{A}~,
~~~~~~~~~~~~{\rm Ric}^{B}{}_{A}~=~R^{B}{}_{ADF}~ g^{FD}~,
\eeq
It is Grassmann odd. The Levi-Civita Ricci tensor \mb{{\rm Ric}} 
vanishes on manifolds which possesses a compatible parity structure.
There is many possibilities of contracting the curvature tensor with the
the metric and the symplectic metric. One way, which actually works
for arbitrary vector bundles, is to contract the two base manifold indices 
with the metric:
\beq
{\rm Rig}~\equiv~ \hf(-1)^{\epsilon_{D}} g^{DA} [\nabla_{A},\nabla_{D}]
~=~ \hf (-1)^{\epsilon_{D}} g^{DA} \partial^{r}_{B}~R^{B}{}_{ADF}
\otimes d\Gamm^{F}~:~TM \to TM~.
\eeq
Much more central for our considerations is the {\em Ricci two-form} $\RR$, 
that only depends on the connection itself. It is by definition the 
curvature two-form for the superdeterminant bundle
\beq
\hf d\Gamm^{A} \RR_{AB}\wedge d\Gamm^{B}
~=~\RR~=~ \hf[\nabla^{\FF} \stackrel{\wedge}{,} \nabla^{\FF}]
 ~=~ -d\Gamm^{A} \wedge d\Gamm^{B}
   (\larrow{\partial^{l}_{B}}\FF_{A})~,
\eeq
so that
\beq
 \RR_{AB}~=~(\larrow{\partial^{l}_{A}}\FF_{B})(-1)^{\epsilon_{B}}
  - (-1)^{(\epsilon_{A}+1)\epsilon_{B}} 
(\larrow{\partial^{l}_{B}}\FF_{A}) 
~=~  (-1)^{(\epsilon_{A}+1)(\epsilon_{B}+1)} \RR_{BA} ~.
\eeq
It could equivally well be defined as the contraction of the bundle indices
of the curvature tensor $R$:
\beq
  \RR~=~{\rm str}R~,~~~~~~~~~ 
(-1)^{\epsilon_{D}} \RR_{AD}
~=~(-1)^{\epsilon_{B}(\epsilon_{A}+\epsilon_{D}+1)} R^{B}{}_{ADB}~.
\eeq
The Ricci two-form \mb{\RR} is closed because of the Bianchi identity,
so it defines a cohomology class \mb{[\RR]}, which up to a normalization
constant is the first Chern class.
On a Ricci-form-flat manifold we can locally
find a bosonic function \mb{\ln\rho} such that
\beq
     \FF_{A}~=~ (\larrow{\partial^{l}_{A}}\ln\rho)~.
\eeq
The function \mb{\rho} transforms as a volume density.
The \mb{\rho} is determined up to a multiplicative constant.

\subsection{Ricci Bracket}

\noi
We can introduce an even Ricci bracket
\beq
(F,G)_{(2)}~=~[[\Delta^2,F],G]1
 ~=~\frac{1}{4}( F \papar{\Gamm^{A}} ) \RR^{AB} (\papal{\Gamm^{A}}G)
~=~(-1)^{\epsilon_{F}\epsilon_{G}}(G,F)_{(2)} ~,
\label{deeahr}
\eeq
where \mb{F,G \in C^{\infty}(M)} are functions. 
Here we have introduce a contravariant version of the Ricci-form tensor
\beq
\RR^{AD}~=~E^{AB} \RR_{BC} E^{CD} 
~=~(-1)^{\epsilon_{A}\epsilon_{B}} \RR^{DA} ~.
\eeq
The Ricci-bracket satisfies a Jacobi identity
\beq
   \sum_{{\rm cycl.}~F,G,H} (-1)^{\epsilon_{F}\epsilon_{H}} 
((F,G)_{(2)},H)_{(2)}~=~0~.
\eeq
This of course reflects the Bianchi identity.
Moreover, the Ricci-bracket has the Poisson property
\beq
  (F,GH)_{(2)}~=~(F,G)_{(2)}H+(F,H)_{(2)}G(-1)^{\epsilon_{G}\epsilon_{H}}~.
\eeq
The relations with the anti-symplectic structure is as 
follows \cite{beringalg}
\bea
 [\Delta,\Delta^2]&=&0~,  \cr
\Delta(F,G)_{(2)}-(\Delta F,G)_{(2)}-(-1)^{\epsilon_{F}}(F,\Delta G)_{(2)}
&=&(-1)^{\epsilon_{F}}\Delta^{2}(F,G) \cr
&&-(-1)^{\epsilon_{F}}(\Delta^{2}F,G)-(-1)^{\epsilon_{F}}(F,\Delta^{2}G)~,\cr
\sum_{{\rm cycl.}~F,G,H} (-1)^{\epsilon_{F}(\epsilon_{H}+1)} ((F,G),H)_{(2)}
&=& \sum_{{\rm cycl.}~F,G,H}(-1)^{\epsilon_{F}(\epsilon_{H}+1)+\epsilon_{G}} 
((F,G)_{(2)},H)~.\cr
&&
\eea

\vs
\noi
We collect the following equivalent conditions in case of a
torsion-free connection:
{\em \begin{enumerate}
\item
The connection \mb{\nabla} is Ricci-form-flat.
\item
The superdeterminant connection \mb{\nabla^{F}} is flat.
\item
\mb{(\larrow{\partial^{l}_{B}}F_{A})
=(-1)^{\epsilon_{A}\epsilon_{B}}
(\larrow{\partial^{l}_{A}}F_{B} )}~.
\item
There exists locally a volume density $\rho$ such that 
\mb{ F_{A}= (\larrow{\partial^{l}_{A}}\ln\rho)}.
\item
$\Delta^2$ is a linear operator:
\beq
   \Delta^2(FG)~=~\Delta^2(F)~G~+~F~\Delta^2(G)~.
\eeq
\item
Leibnitz rule for $\Delta$ and the antibracket: 
\beq
 \Delta(F,G)~=~ (\Delta F,G)-(-1)^{\epsilon_{F}}(F,\Delta G)~.
\eeq
\end{enumerate}} 

%

\subsection{Almost Tensors}

\noi
Let us define differentiation operators
\beq
\tilde{\partial}_{A}~=~(P^T)_{A}{}^{B}~\partial_{A}~,~~~~~~
(\partial^{\sigma})_{A}~=~(P_{\sigma})_{A}{}^{B}~\partial_{A}~.
\eeq
A contravariant almost vector \mb{\tilde{X}^{A}} transforms as
\beq
   \tilde{X}'^{B}~=~\tilde{X}^{A}~
( \larrow{\tilde{\partial}_{A}}\Gamm'^{B})~,
\eeq
under change of coordinates \mb{\Gamm^{A} \to \Gamm'^{A}(\Gamm)}. 
An $(r,s)$ almost tensor is the obvious generalization. Almost tensors 
behaves as tensors under parity preserving transformations.
With this notation the Levi-Civita $\Delta$ can neatly be written as
\beq
\Delta~=~\hf (-1)^{\epsilon_{A}}g^{AB} \larrow{\partial^{l}_{B}}
(P^T)_{A}{}^{C}\larrow{ \partial^{l}_{C}}
~=~- \hf (-1)^{\epsilon_{A}}E^{AB} 
\larrow{\tilde{\partial}^{l}_{B}}
\larrow{\tilde{\partial}^{l}_{A}}
~=~ \hf (-1)^{\epsilon_{A}}E^{AB} 
\larrow{\partial^{l}_{B}}
\larrow{\partial^{l}_{A}}~.
\label{lcnice}
\eeq
The last equality in general only holds for $P$-adapted coordinates.

\subsection{Determinant Density}
\label{determinantdensitysec}

\noi
In this Subsection we consider only the Levi-Civita connection.
By use of the Jacobi identity, one can write the contracted Christoffel
symbol $F_{A}$ as a sum of two contributions:
\beq
   F_{A} ~=~(P^T)_{A}{}^{B} \tilde{F}_{B} + p_{A}~,
\eeq
where 
\bea
  \tilde{F}_{A}\equiv  P^{B}{}_{C} \Gamma^{C}{}_{BA}(-1)^{\epsilon_{A}}
&=& - \hf (-1)^{\epsilon_{B}}
(\larrow{\partial^{l}_{A}}E_{BC}) g^{CB}
~=~- \hf (-1)^{\epsilon_{B}}
(\larrow{\partial^{l}_{A}}g^{BC}) E_{CB} \cr
 &=& - \hf (-1)^{\epsilon_{B}}
(\larrow{\partial^{l}_{A}}E^{BC}) g_{CB}
~=~- \hf (-1)^{\epsilon_{B}}
(\larrow{\partial^{l}_{A}}g_{BC}) E^{CB}~,
\eea
and
\beq
(-1)^{\epsilon_{A}}p_{A}
~\equiv~ (-1)^{\epsilon_{B}}(P^{T})_{B}{}^{C}
(\larrow{\partial^{l}_{C}}P^{B}{}_{A}) 
~=~- (-1)^{\epsilon_{B}}(\larrow{\partial^{l}_{B}}P^{B}{}_{C})
 P^{C}{}_{A} ~.
\eeq
\mb{\tilde{F}_{A}} is zero in Darboux coordinates. 
It transforms under coordinate transformations 
\mb{\Gamm^{A}} \mb{\to} \mb{\Gamm'^{A}(\Gamm)} as a
connection-like field:
\beq
   \tilde{F}_{A}~=~(\larrow{\partial^{l}_{A}}  \Gamm^{A'}) \tilde{F}_{A'}
 + (\larrow{ \partial^{l}_{A}}   \Gamm^{B'}\rarrow{\partial^{r}_{B}}) 
(\Gamm^{B}\rarrow{\partial^{r}_{A'}}) P^{A'}{}_{B'}(-1)^{\epsilon_{B'}} ~.
\label{bvad}
\eeq
Let us now impose that $P$ is a parity structure.
Note that among $P$-adapted coordinates \eq{bvad} reduces to
\beq
   \tilde{F}_{A}~=~(\larrow{\partial^{l}_{A}}  \Gamm^{A'}) \tilde{F}_{A'}
  + (\larrow{ \partial^{l}_{A}} \ln {\rm det}
( \larrow{  \partial^{l}_{\cdot}} \Gamm'^{\cdot} ) ) ~.
\eeq
So in that case \mb{\nabla^{\tilde{F}}=d+\tilde{F}}, 
where \mb{\tilde{F}=d\Gamm^{A}\tilde{F}_{A}} is just a
connection in the determinant bundle over the manifold.
The corresponding curvature two-form is
\beq 
R^{\tilde{F}}~=~\hf[\nabla^{\tilde{F}} \stackrel{\wedge}{,}
 \nabla^{\tilde{F}}]
~=~- d\Gamm^{A} \wedge d\Gamm^{B}
 ( \larrow{ {\partial}^{l}_{B}} \tilde{F}_{A})~\equiv~0~.
\eeq
The second equality is true also in non-$P$-adapted coordinates.
\mb{R^{\tilde{F}}} is identically zero because \mb{\tilde{F}_{A}} vanishes in 
Darboux coordinates. So the connection \mb{\nabla^{\tilde{F}}} is flat.
Hence, there exists locally a density-like function \mb{\tilde{\rho}}, 
such that
\beq
 \tilde{F}_{A}~=~( \larrow{ \partial^{l}_{A}} \ln \tilde{\rho})~.
\eeq
\mb{\tilde{\rho}} is uniquely determined up to a multiplicative constant.
Moreover, \mb{\tilde{\rho}} is a constant in Darboux coordinates.
Notice that in $P$-adapted coordinates \mb{p_{A}=0} vanishes, so that
in these coordinates
\beq
  F_{A} ~=~(P^T)_{A}{}^{B} \tilde{F}_{B}~,~~~~~~~~~~~~ 
\tilde{F}_{A}~=~( \larrow{ \partial^{l}_{A}} \ln \tilde{\rho})~.
\label{lcverynice}
\eeq
If there are two $P$-compatible anti-symplectic structures \mb{E^{(1)}}
and \mb{E^{(2)}}, then the difference of the connection fields 
\mb{\tilde{F}^{(2)}-\tilde{F}^{(1)}} is a global
closed Grassmann-even one-form. Locally, 
\beq
  \tilde{F}^{(2)}_{A}-\tilde{F}^{(1)}_{A}
~=~( \larrow{ \partial^{l}_{A}} \ln 
\frac{\tilde{\rho}^{(2)}}{\tilde{\rho}^{(1)}})~.
\eeq
We shall see in the next Section that for a so-called odd
K\"{a}hler manifold, many of the above local statements holds
globally.

\vs
\noi
Let us note for later that we can rewrite the divergence as
\beq
  {\rm div}(X)
~=~ ( (-1)^{\epsilon_{A}} (P^{T})_{A}{}^{B}\papal{\Gamm^{B}} P^{A}{}_{C} 
 + (-1)^{\epsilon_{C}}(P^{T})_{C}{}^{B}\tilde{F}_{B}) X^{C}~.
\eeq
By previous definition we get 
\beq
  p_{A}~=~(\larrow{\partial^{l}_{A}}\ln\rho) 
-  (\larrow{\tilde{\partial}^{l}_{A}}\ln\tilde{\rho})~.
\eeq

\setcounter{equation}{0}
\section{Odd K\"{a}hler Manifolds}
\label{lonekaehlermann}

\subsection{Definition}

\noi
Recall from Section~\ref{parity} that an {\em odd K\"{a}hler manifold} 
\mb{(M,P,E)} is a \mb{(\nnn | \nnn)} supermanifold $M$ with a non-degenerate 
anti-symplectic structure $E$ and a parity structure $P$ that are mutual 
compatible. The corresponding metric $g$ is called an 
{\em odd K\"{a}hler metric}. An {\em odd K\"{a}hler manifold} \mb{(M,P)} 
is a \mb{(\nnn | \nnn)} supermanifold $M$ with a parity structure $P$, that
{\em admits} an odd K\"{a}hler metric.

\vs
\noi
{\em In the next two Sections we consider an odd K\"{a}hler manifold 
\mb{(M,P,E)}.} 
As a consequence of the definition the Levi-Ceivita connection \mb{\nabla} 
preserves the parity structure $P$ and it is a symplectic connection, 
cf.\ Subsection~\ref{symplcon}. The holonomy group is 
\mb{G \subseteq GL(\nnn)}. This should be compared with the holonomy 
group \mb{G \subseteq U(\nnn)} of the usual K\"{a}hler manifolds. 
Also note that the Lie-bracket between a 
``holomorphic'' and an ``anti-holomorphic'' vector field vanishes. In  a 
$P$-adapted  coordinate system the algebra of vector fields
\beq
   [\papal{\Gamm^{A}},\papal{\Gamm^{B}}]
~=~c_{AB}{}^{C}(\Gamm)\papal{\Gamm^{C}}
\eeq
reduces into two independent algebras, \ie the only non-vanishing components 
of the structure functions \mb{c_{AB}{}^{C}(\Gamm)} are the boso-bose-bose 
and the fermi-fermi-fermi components.

\subsection{K\"{a}hler Potential}

\noi
Note that although a K\"{a}hler potential $K$ may only be locally defined, 
the formula \eq{ekek} ($=$ \eq{ekekek} below) still holds for 
{\em arbitrary} $P$-adapted coordinate systems. In other words, it makes
 sense to view $K$ as a 
local odd scalar function. Moreover, the odd Laplacian on $K$ is equal to
half the dimension of the manifold:
\beq
  (\Delta K)~=~\nnn~.
\eeq

\subsection{Canonical Determinant density}

\noi
There is a {\em globally} defined determinant density \mb{\tilde{\rho}},
cf.\ Subsection~\ref{determinantdensitysec}. In $P$-adapted 
coordinate systems it is defined as the determinant
\beq
   \tilde{\rho}~=~ {\rm det}(E_{\bar{\alpha}\alpha})
\label{kkrho}
\eeq 
of the purely bosonic matrix
\beq
   E_{\bar{\alpha}\alpha}
~=~ (\papal{\theta^{\bar{\alpha}}} K \papar{x^{\alpha} }) ~. 
\label{ekekek}
\eeq
Note that \mb{\tilde{\rho}} within $P$-adapted coordinate systems behaves as 
a determinant density. It is straightforward to see that \mb{\tilde{\rho}} 
up to a multiplicative constant is equal to the locally defined 
density-like object of Section~\ref{determinantdensitysec}. Thus the 
locally defined \mb{\tilde{\rho}} of Section~\ref{determinantdensitysec} 
can be patched together to yield a globally defined density-like object. 
Let us (as always in this paper) assume that the supermanifold $M$ is 
orientable, so that there exists an atlas of $P$-adapted charts such that 
\mb{{\rm body}(\tilde{\rho})>0} everywhere. 
(This should be compared with the situation for (bosonic) complex manifolds. 
These are automatically orientable.) We also concluded in 
Section~\ref{determinantdensitysec}, cf.\  \eq{lcverynice}, that
\beq
   F_{A}~=~(P^{T})_{A}{}^{B}\tilde{F}_{B}~,~~~~~~~~~
\tilde{F}_{B}~=~(\larrow{\partial^{l}_{B}} \ln \tilde{\rho})~.
\label{nicesimplification}
\eeq
{}From \eq{nicesimplification}
follows directly that the Ricci-form \mb{\RR} is a \mb{(1,1)}-form
wrt.\ the Dolbeault bi-complex.
\beq
 \RR ~=~2( \bar{\partial}\wedge \partial \ln \tilde{\rho})~,
\eeq
However, \mb{\RR} is not necessary exact, because \mb{\ln \tilde{\rho}} is
not a scalar object. Note that if \mb{E'} is another $P$-compatible
anti-symplectic structure, the corresponding Ricci-form \mb{\RR'} belongs 
to the same cohomology class 
\beq
 \RR'-\RR~=~2( \bar{\partial} \wedge \partial
\ln \frac{\tilde{\rho}'}{\tilde{\rho}})~.
\eeq
Thus the cohomology class \mb{[\RR]} of the Ricci-form, \ie the first Chern 
class up to a proportionality factor, is independent of the 
anti-symplectic structure.  So if $M$ admits a Ricci-form-flat 
anti-symplectic structure, then the first Chern class vanishes.

\setcounter{equation}{0}
\section{Odd Calabi-Yau Manifolds}

\subsection{Definition}

\noi
Let us define\footnote{Often in the Literature compactness is included in 
the definition of a Calabi-Yau manifold, but we shall not do so.} an 
{\em odd Calabi-Yau manifold} \mb{(M,P)} to be a K\"{a}hler manifold 
\mb{(M,P)} with a vanishing first Chern class.

\vs
\noi
If we naively should translate Calabi's 
conjecture \cite{calabiyau} into the odd case, we are asking whether there 
exists a $P$-compatible  anti-symplectic two-form \mb{E'} in the same 
K\"{a}hler class \mb{[E]}, such that the corresponding Levi-Cevita 
Ricci-form \mb{R'\equiv 0} vanishes identically?
Calabi managed (in the case of usual K\"{a}hler manifolds) to prove the 
uniqueness of the Ricci-form-flat K\"{a}hler $2$-form \mb{[\omega']} 
using the very powerful maximum-modulus principle of complex analysis. 
But we don't have such a principle in the odd case. And in fact we shall 
soon see that uniqueness does not hold in the odd case. The existence
question (the analogue of Calabi's conjecture, \ie Yau's 
Theorem \cite{calabiyau}) is an interesting question, that we however 
shall not address in this paper.
 
\vs
\noi
Let us mention that an important consequence of the Ricci-form flatness 
condition is that the holonomy group is \mb{G \subseteq SL(\nnn)}.
This should be compared with the holonomy group \mb{G \subseteq SU(\nnn)}
for the usual Calabi-Yau manifold.

\subsection{Holomorphic $\nnn$-form}

\noi
Let us {\em assume} that the Ricci-form-tensor is of the form 
\mb{\RR=2( \bar{\partial} \wedge \partial\ln s)}, where \mb{s=s(x,\theta)}
is a global scalar function with \mb{{\rm body}(s)>0}. 
Locally, this is of course true. (In Subsection~\ref{dollydagger}
we gave a sufficient condition for this to be true globally.
However, the assumptions given there are pretty restrictive, so we prefer
just to assume the global existence. After all, the most interesting
case is \mb{s\equiv 1}.) So
\beq
 (\papal{x^{\alpha}} \papal{\theta^{\bar{\alpha}} } 
\ln \frac{\tilde{\rho}}{s})~=~0~.
\eeq
The full solution is that \mb{\tilde{\rho} s^{-1}} factorizes
\beq
   \tilde{\rho}(x,\theta)~s(x,\theta)^{-1}
 ~=~\rho_{+}(x) ~\rho_{-}(\theta)^{-1}~,
\eeq
where \mb{\rho_{+}(x)} and \mb{\rho_{-}(\theta)} are invertible
Grassmann-even functions. Clearly, they are uniquely determined
up to a multiplicative constant. If the first \v{C}ech cohomology class 
\mb{\check{H}^{1}_{\rm const} (M,\R)=0} vanishes, the 
\mb{\rho_{+}(x)} and \mb{\rho_{-}(\theta)} are globally defined,
see Subsection~\ref{dollydagger}. Then there exists a globally
defined nowhere-vanishing ``holomorphic'' $\nnn$-form 
\beq
\Omega_{\rm vol}^{(+)}~=~d^{\nnn}x~\rho_{+}
~=~\frac{1}{\nnn!} \Omega^{(+)}_{\alpha_{1}\alpha_{2}\ldots\alpha_{\nnn}}
dx^{\alpha_{1}}\wedge dx^{\alpha_{2}} \wedge\ldots \wedge dx^{\alpha_{\nnn}}~.
\eeq
It is straightforward to derive that 
\beq
  \nabla\Omega_{\rm vol}^{(+)}~=~0 ~~~\Leftrightarrow~~~ (\partial s)~=~0~.
\eeq
So the ``holomorphic'' $\nnn$-form is covariantly preserved iff
the anti-symplectic structure is Ricci-form-flat. If 
\mb{\Omega_{\rm vol}^{(+)'}=d^{\nnn}x~\rho'_{+}} is another 
nowhere-vanishing ``holomorphic'' $\nnn$-form, then 
\mb{s'=\frac{\rho'_{+}}{\rho_{+}}} is a nowhere-vanishing ``holomorphic''
scalar. If both $\nnn$-forms are covariantly preserved, then $s'$ is a 
constant on each connected components of $M$. In particular we have proven:

\vs
\noi
{\bf Proposition}.
{\em Given a connected odd K\"{a}hler manifold $M$ that admits an 
$P$-compatible Ricci-form-flat anti-symplectic structure. A 
nowhere-vanishing, covariantly preserved, ``holomorphic'' $\nnn$-form 
is unique up to a multiplicative constant. If the first \v{C}ech cohomology 
class \mb{\check{H}^{1}_{\rm const} (M,\R)=0} vanishes, then there exists 
a globally defined nowhere-vanishing, covariantly preserved, 
``holomorphic'' $\nnn$-form \mb{\Omega_{\rm vol}^{(+)}}.}

\subsection{Canonical Volume Form}

\noi
Let $M$ be a Ricci-form-flat odd K\"{a}hler manifold.
If the nowhere-vanishing, covariantly preserved, ``holomorphic'' $\nnn$-form 
\mb{\Omega_{\rm vol}^{(+)}} exists, we may define an ``anti-holomorphic''
pendant
\beq
\Omega^{(-)}~=~\frac{1}{\nnn!} 
\Omega^{(-)}_{\bar{\alpha}_{1}\bar{\alpha}_{2}\ldots\bar{\alpha}_{\nnn}}
d\theta^{\bar{\alpha}_{1}}\vee d\theta^{\bar{\alpha}_{2}} \vee
\ldots \vee d\theta^{\bar{\alpha}_{\nnn}}~,~~~~~~~~~
\frac{1}{n!}\Omega^{(-)}_{\bar{\alpha}_{1}\bar{\alpha}_{2}
\ldots\bar{\alpha}_{\nnn}}
~=~\rho_{-}(\theta)^{-1}~=~\frac{\tilde{\rho}(x,\theta)}{\rho_{+}(x)}~.
\eeq
Recall that for Grassmann-odd variables differential forms are {\em not}
directly related to integration theory and volume forms.
However in the case at hand, we may easily construct a volume form
\mb{\Omega_{\rm vol}^{(M)}=d^{2\nnn}\Gamm~\rho}. 
In a local $P$-adapted coordinate system the density \mb{\rho} is simply
\beq
   \rho~=~ \rho_{+}(x)~\rho_{-}(\theta)~.
\label{krho}
\eeq

\subsection{Nilpotency}

\noi
In the case of an odd K\"{a}hler manifold $M$, we can write the square of
the odd Levi-Civita Laplacian in $P$-adapted coordinates as
\beq
 \Delta^{2}~=~ 
\hf \RR^{\bar{\alpha}\alpha}\papal{x^{\alpha}}\papal{\theta^{\bar{\alpha}}}~.
\eeq
This fact follows from \eq{deeahr} and the observation that 
\mb{\Delta^2} has no monomial with less than two differentiations when 
\mb{\Delta^2} is normal-ordered, \ie differential operators ordered to 
the right, cf.\ expression \eq{sonice} or expression \eq{lcnice}.
As a consequence we have:

\vs
\noi
{\bf Theorem.} {\em Let $M$ be an odd K\"{a}hler manifold. $M$ is 
Ricci-form-flat if and only if the odd Levi-Civita Laplacian \mb{\Delta} 
is nilpotent.}

\subsection{Standard Example}

\noi
The standard example of an odd Calabi-Yau manifold is
the cotangent bundle \mb{M=\Omega^{1}(L)} of a bosonic $\nnn$-dimensional
manifold $L$ that satisfies certain extra properties listed below. We 
identify \mb{\theta^{\bar{\alpha}}=dx^{\bar{\alpha}}}. The Grasmann parity 
(\ie the form parity) is here a global parity structure $P$. 
The manifold $L$ should be equipped with a covariant 
non-degenerate two-tensor 
\beq
  g~=~dx^{\alpha}~g_{\alpha\bar{\alpha}}(x) \otimes\theta^{\bar{\alpha}}  
  ~\in~ \Gamma(T^{*}(L)\otimes T^{*}(L))
\eeq
such that
\beq
   (\larrow{\partial^{l}_{\alpha}}g_{\beta\bar{\alpha}})~=~
   (\larrow{\partial^{l}_{\beta}}g_{\alpha\bar{\alpha}})~.
\eeq
So locally we can introduce a K\"{a}hler potential 
\mb{K=\eta_{\bar{\alpha}}(x)\theta^{\bar{\alpha}}},
where 
\beq
(\larrow{\partial^{l}_{\alpha}}\eta_{\bar{\alpha}})
 ~=~g_{\alpha\bar{\alpha}}~.
\eeq
We introduce for convenience the ``transposed'' matrix 
\mb{g_{\bar{\alpha}\alpha}~=~g_{\alpha\bar{\alpha}}}.
The Ricci-form-flat anti-symplectic structure is 
\mb{E=d\theta^{\bar{\alpha}}g_{\bar{\alpha}\alpha}\wedge dx^{\alpha}}.
In the generic case, the Ricci-form-flat structure (already within a given
cohomology class) is not unique. So Calabi's Uniqueness Theorem does
not hold for the odd case.

\subsection{Lagrangian Density}

\noi
Now fix a point \mb{m \in M} in the manifold and consider the two 
Lagrangian surfaces \mb{L_{\pm}} that intersect $m$. 
Contrary to even symplectic geometry the canonical volume form 
\mb{\Omega_{\rm vol}^{(M)}} actually induces \cite{schwarz} canonical 
volume forms \mb{\Omega_{\rm vol}^{(\pm)}}
on the Lagrangian surfaces \mb{L_{\pm}}, respectively. 
They are defined via
\bea
  \Omega_{\rm vol}^{(+)}(e_{(1)},\ldots,e_{(\nnn)})
&=&  \sqrt{\Omega_{\rm vol}^{(M)} (e_{(1)},\ldots, e_{(\nnn)},
e_{*}^{(1)},\ldots, e_{*}^{(\nnn)})}~, \cr
  \Omega_{\rm vol}^{(-)}(e_{(\bar{1})},\ldots,e_{(\bar{\nnn})})
&=& \sqrt{ \Omega_{\rm vol}^{(M)} (e_{(\bar{1})},\ldots, e_{(\bar{\nnn})},
e_{*}^{(\bar{1})},\ldots, e_{*}^{(\bar{\nnn})})}
\eea
Here \mb{(e_{(\alpha)})_{\alpha\!=\!1,\ldots, \nnn}} and 
\mb{(e_{(\bar{\alpha})})_{\bar{\alpha}\!=\!\bar{1},\ldots, \bar{\nnn}}} 
denotes a basis for the tangent spaces \mb{TL_{\pm}}, respectively, 
and $*$ denotes the odd parity conjugation, cf.\ Section~\ref{oddparityconj}.
{}From the formulas
\beq
   \partial^{l,\alpha}_{*}
~=~E^{\alpha\bar{\alpha}}\partial^{l}_{\bar{\alpha}}~,~~~~~~~~~~~~~~~~
   \partial^{l,\bar{\alpha}}_{*}
~=~E^{\bar{\alpha}\alpha}\partial^{l}_{\alpha}~,
\eeq
we conclude (up to an inessential over-all sign convention) that
\beq
\Omega_{\rm vol}^{(+)}~=~d^{\nnn}x~\rho_{+}~,~~~~~~~~~~~~~~~~
\Omega_{\rm vol}^{(-)}~=~d^{\nnn}\theta~\rho_{-}~.
\eeq
In other words, the bosonic Lagrangian volume density 
\mb{\Omega_{\rm vol}^{(+)}} is precisely the previously mentioned
``holomorphic'' $\nnn$-form.

\subsection{Odd Symplectic Potential}

\noi
There are three natural choices of a symplectic potential.
Perhaps the most symmetric choice is \mb{\vartheta=\vartheta'}, where 
\beq
\begin{array}{rclcrcl}
  \vartheta'&=& \vartheta'_{A} d\Gamm^{A}&~~~~~~~~~~~~~~~~~&
  \vartheta'_{A}&=&\hf (P^T)_{A}{}^{B} (\papal{\Gamm^{B} } K)~, \cr \cr
    \vartheta'_{\alpha}&=& \hf (\papal{x^{\alpha}} K) &&
\vartheta'_{\bar{\alpha}}&=&  -\hf(\papal{\theta^{\bar{\alpha}}} K) ~.
\end{array}
\eeq
Note that the first two formulas are covariant. They hold in arbitrary 
coordinates. So if the K\"{a}hler potential $K$ is globally defined, so is 
the odd symplectic potential \mb{\vartheta'}. A second choice is
\beq
\begin{array}{rclcrcl}
  \vartheta&=& \vartheta_{A} d\Gamm^{A} &~~~~~~~~~~~~~~~~~~&
  \vartheta_{A}&=&(P_{+}^T)_{A}{}^{B} (\papal{\Gamm^{B} } K)~, \cr \cr
    \vartheta_{\alpha}&=&(\papal{x^{\alpha}} K)  &&
\vartheta_{\bar{\alpha}}&=&0~.
\end{array}
\eeq
We can change coordinates 
\mb{(x^{\alpha};\theta^{\bar{\alpha}}) \to (x^{\alpha};\vartheta_{\alpha})} 
to Darboux coordinates \mb{(x^{\alpha};\vartheta_{\alpha})},
\beq
  (x^{\alpha},x^{\beta})~=~0~,~~~~~~
  (x^{\alpha},\vartheta_{\beta}) ~=~\delta^{\alpha}_{\beta}~,~~~~~~~
 (\vartheta_{\alpha} ,\vartheta_{\beta}) ~=~ 0~.
\eeq
The super Jacobian of the coordinate transformation is 
\mb{J={\rm det}(E_{\bar{\alpha}\alpha})^{-1}=\rho_{-}^{}\rho_{+}^{-1}}, 
so the canonical volume density in these Darboux variables reads
\beq
   \rho~=~\rho_{+}^{2}~.
\eeq
Hence the $\Delta$ operator becomes
\beq
   \Delta ~=~E^{\alpha\bar{\alpha}} \papal{\theta^{\bar{\alpha}}} 
(\papal{x^{\alpha}})\sokkel{\theta}
 ~=~\papal{\vartheta_{\alpha}} 
\rho_{+}^{-1}(\papal{x^{\alpha}})\sokkel{\vartheta}\rho_{+}~.
\eeq
The new coordinates \mb{(x^{\alpha};\vartheta_{\alpha})} does {\em not} 
adapt the $P$-structure. In a Grassmann \mb{2\times2} block representation
the parity structure reads
\beq
   P~=~ \twobytwo{{\bf 1}}{0}{2{\bf K}}{-{\bf 1}}~,
\eeq
where \mb{{\bf K}} denote the \mb{\nnn\times\nnn} matrix with entries
\beq
{\bf K}_{\alpha\beta} ~=~
(\papal{x^{\alpha}} K\papar{x^{\beta}})\sokkel{\theta}~.
\eeq
Despite this fact, they are related to a Fourier transform, see next 
Section. Moreover, in a mixed notation, where we use {\em both} coordinate 
systems \mb{(x^{\alpha};\theta^{\bar{\alpha}})} and 
\mb{(x^{\alpha};\vartheta_{\alpha})}, the odd Laplacian \mb{\Delta} operator 
acquires a Darboux-like form
\beq
  \Delta ~=~ \papal{\vartheta_{\alpha}} 
(\papal{x^{\alpha}})\sokkel{\theta}~,~~~~~~~~
(\papal{x^{\alpha}})\sokkel{\theta}~=~(\papal{x^{\alpha}})\sokkel{\vartheta}
+(\papal{x^{\alpha}} K\papar{x^{\beta}})\sokkel{\theta}
\papal{\vartheta_{\beta}}~.
\eeq
Let us for completeness list the third natural choice \mb{\vartheta=\yy} 
\beq
\begin{array}{rclcrcl}
  \yy&=& \yy_{A} d\Gamm^{A} &~~~~~~~~~~~~~~~~~~&
  \yy_{A}&=& (P_{-}^T)_{A}{}^{B} (\papal{\Gamm^{B} } K)~, \cr \cr
    \yy_{{\alpha}}&=&0  &&
\yy_{\bar{\alpha}}&=& -(\papal{\theta^{\bar{\alpha}}} K).
\end{array}
\eeq
We now choose Darboux coordinates 
\mb{(\yy_{\bar{\alpha}};\theta^{\bar{\alpha}})}
\beq
  (\yy_{\bar{\alpha}},\yy_{\bar{\beta}})~=~0~,~~~~~~
  (\yy_{\bar{\alpha}},\theta^{\bar{\beta}})
 ~=~\delta_{\bar{\alpha}}^{\bar{\beta}}~,~~~~~~~
 (\theta^{\bar{\alpha}},\theta^{\bar{\beta}}) ~=~ 0~.
\eeq
The super Jacobian of the coordinate transformation 
\mb{(x^{\alpha};\theta^{\bar{\alpha}}) \to 
(\yy_{\bar{\alpha}};\theta^{\bar{\alpha}})} is 
\mb{J={\rm det}(E_{\bar{\alpha}\alpha})=\rho^{}_{+}\rho_{-}^{-1}}, so the 
canonical volume density is \mb{\rho=\rho_{-}^2}.

\subsection{Fourier Transform}

\noi
Let there be given a $P$-adapted bosonic \mb{(\nnn |0)} Lagrangian surface 
\mb{L} in \mb{M} and a globally defined K\"{a}hler potential $K$. We 
equip \mb{L} with the canonical volume density \mb{\rho_{L}=\rho_{+}} on 
\mb{L}.
In a $P$-adapted coordinate system, we may write
\beq 
L~=~
\{ (x^{\alpha};\theta^{\bar{\alpha}}) |  
\theta^{\bar{\alpha}}=\theta_{(0)}^{\bar{\alpha}} \}~.
\eeq 
We may assume that the local $P$-adapted charts in the atlas are of the 
box-type \mb{U \times \R^{\nnn}_{a}}, where \mb{U\subseteq \R^{\nnn}_{c}} 
and \mb{\R_{c}} (\mb{\R_{a}}) denotes the set of Grassmann-even  
(Grassmann-odd) real supernumbers, respectively, see \cite{berezindewitt}.
In other words, we may assume that the $P$-adapted fermionic 
\mb{(0|\nnn )} Lagrangian surfaces are covered by a single chart.
\beq 
L_{-}(x_{(0)})~=~
\{ (x^{\alpha};\theta^{\bar{\alpha}}) |  x^{\alpha}=x_{(0)}^{\alpha} \}~.
\eeq
We now define a Fourier transform 
\mb{\wedge: C^{\infty}(M) \to \Omega^{\bullet}(L)}
from the space of functions \mb{f(x,\theta)} on $M$ to the exterior algebra 
of forms \mb{\hat{f}(x,c)} on the Lagrangian surface \mb{L}.
\bea
   \hat{f}(x,c)&=&  \int_{L_{-}(x)} \! \! \exp\left[ c^{\alpha}(
\vartheta_{\alpha}(x,\theta)
-\vartheta_{\alpha}(x,\theta_{(0)})
)\right] ~f(x,\theta)~\Omega_{\rm vol}^{(-)}(\theta)  \cr
&=&  \int_{L_{-}(x)} \! \! \exp\left[ c^{\alpha}(
\vartheta_{\alpha}(x,\bar{\theta}+\theta_{(0)})
-\vartheta_{\alpha}(x,\theta_{(0)})
)\right] ~f(x,\bar{\theta}+\theta_{(0)})~
d^{\nnn}\bar{\theta}~\rho_{-}(\bar{\theta}+\theta_{(0)}) \cr
&=& \rho_{+}(x)\int \! \! e^{c^{\alpha}\bar{\vartheta}_{\alpha}}
f(x,\bar{\theta}(x,\bar{\vartheta})+\theta_{(0)})~d^{\nnn} \bar{\vartheta}~.
\eea
Note that the Fourier transform acts from the right.
Here the Grassmann-odd variables \mb{c^{\alpha}} play the r\^{o}le of 
the natural basis of one-forms \mb{dx^{\alpha}}, \ie they
transform in the same way with the form-degree
replaced by the Grassmann-degree. In the second equality, we substituted
\beq
     \bar{\vartheta}_{\bar{\alpha}}~=~
(\papal{x^{\alpha}} K)\sokkel{\theta}(x,\bar{\theta}+\theta_{(0)})
-(\papal{x^{\alpha}} K)\sokkel{\theta}(x,\theta_{(0)})~.
\eeq 
Clearly \mb{\bar{\theta} \to \bar{\vartheta}} is a bijection, whose 
inverse we denote by \mb{\bar{\theta}=\bar{\theta}(x,\bar{\vartheta})}.
We have
\bea
\widehat{(\frac{\partial^{l}}{{\partial \vartheta_{\alpha}}}f)}
&=&c^{\alpha}\hat{f}~, \cr
\wedge\left(\rho_{+}^{-1}(\frac{\partial^{l}}{{\partial x^{\alpha}}})
\sokkel{\vartheta}
\rho_{+} f\right)
 &=&  (\frac{\partial^{l}}{{\partial x^{\alpha}}} \hat{f})~, \cr
\wedge\left((\vartheta_{\alpha}- \vartheta_{\alpha}(\theta_{(0)}))f\right)
&=&(\frac{\partial^{l}}{{\partial c^{\alpha}}}\hat{f})~.
\eea
In particular, the Fourier transform of the odd Laplacian is the
exterior derivative \cite{witten}.
\beq
  \widehat{\Delta f}~=~d \hat{f}~,~~~~~~~~~~~
d~=~ c^{\alpha}\papal{x^{\alpha}}~. 
\eeq
The inverse Fourier transform reads
\beq
f(x,\theta)~=~ \rho_{+}(x)^{-1} \int \! \! 
\exp\left[(\vartheta_{\alpha}(x,\theta)-\vartheta_{\alpha}(x,\theta_{(0)}))
c^{\alpha}\right]\hat{f}(x,c)~d^{\nnn}c~.
\eeq
Note that the fermionic top-monomial 
\beq
 \prod_{\bar{\alpha}}
(\theta^{\bar{\alpha}}-\theta_{(0)}^{\bar{\alpha}} )~~~~~\propto~~~~~
 \frac{ \prod_{\bar{\alpha}}
(\theta^{\bar{\alpha}}-\theta_{(0)}^{\bar{\alpha}} ) 
}{\rho_{-}(\theta)} ~~~~~
\stackrel{\wedge}{\mapsto}~~~~~1~,
\eeq
is mapped to a constant by the Fourier transform.
But the Poincar\'{e} Lemma states that the only non-trivial local DeRahm 
cohomology is the constant zero-forms. We therefore have:

\vs
\noi 
{\bf Local Cohomology Theorem.}
{\em Given a Ricci-form-flat  odd K\"{a}hler  manifold $M$. 
Locally, the solutions $f$ to the equation  
\beq
   (\Delta f)  ~=~ 0~.
\eeq
are of the form
\beq
 f ~=~(\Delta\Psi) + c \Theta ~,
\eeq
where $\Psi$ is a function of opposite Grassmann parity,
$c$ is a constant and \mb{\Theta} is the fermionic top-monomial:
\beq
 \Theta ~\equiv~ \prod_{\bar{\alpha}} \theta^{\bar{\alpha}} ~.
\eeq} 

\vs
\noi
Whereas \mb{\Theta} is not a covariant object the one-dimensional 
\mb{\Delta}-cohomology class \mb{\{[c \Theta]| c \in \R \}} is. 
The fact that \mb{\Delta} has non-trival cohomology given by the
fermionic top-monomial has previously been reported in \cite{beringthesis}.


\setcounter{equation}{0}
\section{Vielbein Formulation}
\label{vielbein}

\subsection{Vielbeins}

\noi
In general, there is no canonical way of introducing an almost parity 
structure $P$. We shall see that a vielbein formulation overcome this 
difficulty.

\vs
\noi
We consider an $(\nnn | \nnn)$ supermanifold $M$ equipped with an
anti-symplectic vielbein \mb{e^{a}{}_{A}}, of Grassmann
parity \mb{\epsilon(e^{a}{}_{A})=\epsilon_{a}+\epsilon_{A}},  
\ie a diffeomorphism
\beq
 e~=~\partial^{r}_{a}~ e^{a}{}_{A}\larrow{d\Gamm^{A}}~:~ TM \to TW~.
\eeq
Here ``$\www$-space'' \mb{TW=W} is an anti-symplectic vector space,
with a constant almost Darboux metric \mb{E_{(0)}^{ab}}.
We denote the basis \mb{\partial^{r}_{a}} and dual basis for \mb{d\www^{a}},
both of Grassmann parity \mb{\epsilon_{a}},
although we will not necesseraly give any sense to a $\www^{a}$ coordinate.
The inverse vielbein map is denoted 
\beq
 e^{-1}~=~\partial^{r}_{A}~ e^{A}{}_{a}~
\larrow{d\www^{a}}~:~ TW \to TM~.
\eeq
We have the orthonormality relations
\bea
  e^{a}{}_{A}~ e^{A}{}_{b}&=&\delta^{a}_{b}~,~~~~~~~~~~~~~~~~~~~~~
  e^{A}{}_{a}~ e^{a}{}_{B}~=~\delta^{A}_{B}~, \cr
 [e,e^{-1}]&=&{\rm Id}_{TW}-{\rm Id}_{TM}~,~~~~~~[e,e]~=~0~,~~~~~
[e^{-1},e^{-1}]~=~0~.
\eea
The anti-symplectic structure is given as
\bea
   E^{AB}&=&       e^{A}{}_{a}~E_{(0)}^{ab}~(e^{T})_{b}{}^{B}~,~~~~~~~~~~~~~
   E_{AB}~=~ (e^{T})_{A}{}^{a}~E^{(0)}_{ab}~      e^{b}{}_{B}~, \cr
  &&~~~~~~E~=~-\hf[e \stackrel{\wedge}{,} 
[ e \stackrel{\wedge}{,} E^{(0)}]]~.
\eea
where the supertransposed vielbein 
\mb{e^{T}=~ d\Gamm^{A}~(e^{T})_{A}{}^{a}\larrow{i^{l}_{a}}~
:~ T^{*}W \to T^{*}M} is
\beq
   (e^{T})_{A}{}^{a}
~=~ (-1)^{(\epsilon_{a}+1)\epsilon_{A}} e^{a}{}_{A}~.
\eeq
Let us also introduce the vielbein one-forms
\beq
   e^{a}~=~e^{a}{}_{A}~d\Gamm^{A}~=~d\Gamm^{A}~(e^{T})_{A}{}^{a}~.
\eeq
We will restrict our attention to vielbein formulations where the 
vielbeins \mb{ \partial^{l(z)}_{a}= (e^{T})_{a}{}^{A} \partial^{l}_{A}}
spans a unit volume-cell of $TM$ (up to a sign):
\beq
   {\rm vol}(\partial^{l(z)}_{a})~=~\pm1~,
\eeq
where ``\mb{{\rm vol}}'' denotes the {\em signed} volume of a frame.
This is a non-trivial but very reasonable requirement for a vielbein 
formulation, thereby linking in a natural way the notion of volume 
density and the notion of anti-symplectic metric. 
A canonical volume density is then given as
\beq
   \rho~~=~~{\rm sdet}(e)~{\rm vol}\left(
\partial^{l(z)}_{a}|a\!=\!1,\ldots,2\nnn\right)~~=~~\pm {\rm sdet}(e)~.
\eeq
Let us for simplicity assume that the manifold is orientable,
so that we can treat ``\mb{{\rm vol}}'' as being single-valued.

\subsection{Canonical Almost Parity Structure}

\noi
The manifold possesses a {\em canonical almost parity structure} $P$,
\beq
  P^{A}{}_{B}~=~ e^{A}{}_{a}~(-1)^{\epsilon_{a}}~ e^{a}{}_{B}~.
\eeq
$P$ is compatible with the anti-symplectic structure $E$ in the sense of
\eq{epep}. In $P$-adapted coordinates, the vielbein \mb{e^{a}{}_{A}} 
acquires a block-diagonal form
\beq
  e^{a}{}_{A}~ \neq ~0 ~~~ \Rightarrow~~~ \epsilon_{a}~=~\epsilon_{A}~.
\eeq
Moreover, the vielbein 
\mb{ \partial^{l(z)}_{a}= (e^{T})_{a}{}^{A} \partial^{l}_{A}} separates 
into ``holomorphic'' and ``anti-holomorphic'' directions for the canonical 
$P$-structure. This can be seen by noting that the vielbein is the 
diagonalizing transformation in the tangent space \mb{TM} for the almost 
parity structure $P$. A coordinate system is called a {\em Grassmann 
preserving coordinate system}, iff all the non-vanishing entries of the 
vielbein is bosonic. Grassmann preserving coordinates are therefore the 
same as $P$-adapted coordinates. The manifold $M$ is an odd pre-K\"{a}hler 
manifold wrt.\ the canonical almost parity structure $P$, if it admits an 
atlas of Grassmann preserving coordinates.

\vs
\noi
We may introduce the ``flat'' bosonic (fermionic) volume factor
\beq
\rho^{(0)}_{\sigma}~=~{\rm vol}(\partial^{l(z)}_{a_{\sigma}})~,~~~~~~~~~
\sigma~=~\pm1~,
\eeq
spanned by the $\nnn$ bosonic (fermionic) vielbeins 
\mb{\partial^{l(z)}_{a}}, \mb{\epsilon_{a}=0} (\mb{\epsilon_{a}=1}), 
respectively.
A vielbein is called {\em special}, iff
\beq
{\rm vol}(\partial^{l(z)}_{a_{+}})~{\rm vol}(\partial^{l(z)}_{a_{-}})
~=~{\rm vol}(\partial^{l(z)}_{a})~.
\eeq
For $P$-adapted coordinates the canonical volume form factorizes.
In case of a special vielbein we may write:
\beq
  \rho~=~\rho_{+}~\rho_{-}~,~~~~~~~~~~~~~
 \rho_{\pm}~=~ {\rm sdet}(e^{a_{\pm}}{}_{A_{\pm}})~\rho^{(0)}_{\pm}
~=~ {\rm det}(e^{a_{\pm}}{}_{A_{\pm}})^{\pm 1}~\rho^{(0)}_{\pm}~.
\eeq

\subsection{Gauge Group}

\noi
There is a structure-preserving gauge group acting on the flat indices 
of the vielbeins. In other words, the group is a subgroup of \mb{GL(TW)}.
(We shall be more explicit about the group below).  A group element 
\beq 
       \Lambda ~=~ \partial^{r}_{a}~\Lambda^{a}{}_{b}\larrow{d\www^{b}}~:~
TW \to TW
\eeq
acts on the vielbein as
\beq
\begin{array}{rcccl}
    e^{a}{}_{A} & \to & (\Lambda.e)^{a}{}_{A}
& = & \Lambda^{a}{}_{b}~e^{b}{}_{A}~, \\
e^{A}{}_{a} ~=~  (e^{-1})^{A}{}_{a} 
& \to & ((\Lambda.e)^{-1})^{A}{}_{a}
 & = & e^{A}{}_{b}~(\Lambda^{-1})^{b}{}_{a}~. 
\end{array}
\eeq
The group action reflects the gauge freedom in choice of vielbein.
The above requirement that the vielbein should represent the volume 
element reduces the gauge group to the subgroup $G$ whose elements
has superdeterminant \mb{\pm1}. 
In the light of the canonical almost parity structure $P$,
we could just as well define the vielbein map 
\mb{e:TM \to TW} in terms of the canonical odd metrics
\beq 
  g_{AB}~=~ (e^{T})_{A}{}^{a}~g^{(0)}_{ab}~e^{b}{}_{B}~,~~~~~~~~~~~
g~=~-\hf[e \stackrel{\vee}{,} [ e \stackrel{\vee}{,} g^{(0)}]]~.
\eeq
where as usual the odd metric is
\beq 
  g_{AB}~=~E_{AC}~P^{C}{}_{B}~,~~~~~~~~~~~
 g^{(0)}_{ab}~=~E^{(0)}_{ab}~(-1)^{\epsilon_{b}}~.
\eeq
We see that the gauge group \mb{G \subseteq GL(TW)} for the flat indices 
should preserve 1) the anti-symplectic metric, 2) the metric and 
3) the volume up to a sign. 
This means the group elements 
\mb{\Lambda ~\in ~G} satisfy
\beq
E_{(0)}^{ad} ~=~ 
\Lambda^{a}{}_{b}~E_{(0)}^{bc}~(\Lambda^{T})_{c}{}^{d}~,~~~~  
g_{(0)}^{ad} ~=~ 
\Lambda^{a}{}_{b}~g_{(0)}^{bc}~(\Lambda^{T})_{c}{}^{d}~,~~~~  
{\rm sdet}(\Lambda)~=~\pm1.
\eeq
$G$ is the subgroup of the orthosymplectic group whose elements
has  superdeterminant \mb{\pm1}. This is isomorphic to 
\mb{GL(\nnn) \cap {\rm det}^{-1}(\{ \pm1 \})}.
\beq
  G~=~Osp(\nnn | \nnn) \cap {\rm sdet}^{-1}(\{ \pm1 \}) 
~\cong~GL(\nnn) \cap {\rm det}^{-1}(\{ \pm1 \})~.
\eeq
To see this, consider the usual Grassmann $2\times2$ 
block representation of the matrices. Let us fix notation 
\beq
\begin{array}{rclcrcl}
   E_{(0)}^{\cdot\cdot}&=&\twobytwo{0}{\Eta^{-1,T}}{-\Eta^{-1}}{0}~,&&
   E^{(0)}_{\cdot\cdot}&=&\twobytwo{0}{-\Eta}{\Eta^{T}}{0}~, \cr \cr
   g_{(0)}^{\cdot\cdot}&=&\twobytwo{0}{\Eta^{-1,T}}{\Eta^{-1}}{0}~, && 
g^{(0)}_{\cdot\cdot}&=& \twobytwo{0}{\Eta}{\Eta^{T}}{0}~,\cr \cr
   P_{(0)}{}^{\cdot}{}_{\cdot}&=&\twobytwo{{\bf 1}}{0}{0}{-{\bf 1}}~.
\end{array}
\label{verydefoff}
\eeq
The group elements \mb{\Lambda^{\cdot}{}_{\cdot}} are of the form
\beq
  \Lambda^{\cdot}{}_{\cdot}
~=~\twobytwo{\lambda}{0}{0}{\Eta^{-1}\lambda^{-1,T} \Eta}~.
\eeq
We shall only discuss real representations.
Then the superdeterminant is positive
\beq 
0~\leq~{\rm det}(\lambda)^{2}~=~{\rm sdet}(\Lambda)~=~\pm 1~,
\eeq
so the \mb{N\times N} bose-bose block \mb{\lambda} has determinant
\beq
  {\rm det}(\lambda)~=~ \pm 1~.
\eeq
Hence we have a non-rigid \mb{GL(\nnn) \cap {\rm det}^{-1}(\{ \pm1 \}) } 
gauge symmetry in each point. We see that the bosonic and the fermionic
volume factors \mb{\rho^{(0)}_{\sigma}} is preserved up to a sign under 
gauge transformation. And the product 
\mb{\rho^{(0)}_{+}~\rho^{(0)}_{-}} is invariant.

\subsection{Levi-Civita Connection}

\noi
Having a canonical odd metric $g$ at our disposal, we can construct 
the Levi-Civita connection 
\beq
   \nabla^{\Gamma}~=~d+\Gamma~:~TM \times TM \to TM~,
\eeq
where
\beq
\Gamma~=~d\Gamm^{A}~ \partial^{r}_{B}~\Gamma^{B}{}_{AC}~\larrow{d\Gamm^{C}} 
 ~=~e^{a}~ \partial^{r(\Gamm)}_{b}~\Gamma^{b}{}_{ac}~\larrow{e^{c}} ~, 
\eeq
and where
\bea
   \Gamma^{b}{}_{ac}&\equiv& (-1)^{(\epsilon_{A}+\epsilon_{a})\epsilon_{b}}
  (e^T)_{a}{}^{A}~ e^{b}{}_{B}~\Gamma^{B}{}_{AC}~ e^{C}{}_{c} \cr
 &=& -(-1)^{(\epsilon_{a}+1)(\epsilon_{C}+1)}
  e^{b}{}_{B}~\Gamma^{B}{}_{CA}~ e^{A}{}_{a}~ e^{C}{}_{c} ~.
\label{gammaup}
\eea
In the second equality of \eq{gammaup}, we used the symmetry property
\eq{reallynotorsion} between the lower indices of the Levi-Civita 
Christoffel symbols. The upper flat index is lowered with the flat
metric:
\bea
\Gamma^{d}{}_{ab}&=&(-1)^{\epsilon_{a}}g_{(0)}^{dc}~ \Gamma_{c,ab}~,\cr
\Gamma_{c,ab}&=&(e^{T})_{c}{}^{C}~\Gamma_{C,BA}~
e^{A}{}_{a}~e^{B}{}_{b}(-1)^{\epsilon_{a}\epsilon_{B}}~. 
\eea
We stress the fact that the flat Christoffel symbols {\em depend} 
on the choice of the curved coordinate system.
To facilitate writing down the formulas we will introduce short hand
notation for the most common combinations of vielbeins. 
We introduce structure functions \mb{f_{a}{}^{b}{}_{c}}, 
\mb{f_{ab}{}^{c}}, \mb{\gamma_{abc}} and \mb{a_{abc}}, mutually 
related as indicated below: 
\bea
f_{a}{}^{b}{}_{c}&\equiv&
(e^{T})_{a}{}^{A}~(\larrow{\partial^{l}_{A}}e^{b}{}_{D} )~e^{D}{}_{c}
~=~ (\larrow{\partial^{l(\Gamm)}_{a}}((\ln e)^{b}{}_{c}) ~, \cr
f_{ab}{}^{c}&\equiv&(e^{T})_{a}{}^{A}~(\larrow{\partial^{l}_{A}}
(e^{T})_{b}{}^{D} )~(e^{T})_{D}{}^{c} \cr
&=& (\larrow{\partial^{l(\Gamm)}_{a}}((\ln e^{-1,T})_{b}{}^{c})
~=~ -(-1)^{\epsilon_{b}(\epsilon_{c}+1)}f_{a}{}^{c}{}_{b} ~, \cr
\gamma_{abc}&\equiv&(e^{T})_{a}{}^{A}~ (\larrow{\partial^{l}_{A}}
(e^{T})_{D}{}^{d} )~g^{(0)}_{db}~e^{D}{}_{c}
(-1)^{\epsilon_{b}\epsilon_{D}} \cr
&=& (-1)^{\epsilon_{c}(\epsilon_{b}+\epsilon_{d}+1)}
f_{a}{}^{d}{}_{c}~g^{(0)}_{db} 
~=~-(-1)^{\epsilon_{b}\epsilon_{c}} f_{ac}{}^{d}~g^{(0)}_{db} ~, \cr
\aaa_{bac}&\equiv&(-1)^{\epsilon_{a}(\epsilon_{d}+1)}
g^{(0)}_{bd}~f_{a}{}^{d}{}_{c}
~=~(-1)^{\epsilon_{a}\epsilon_{b}}\gamma_{abc}~.
\eea
The Grassmann parity is
\beq
  \epsilon(f_{a}{}^{b}{}_{c})~=~ \epsilon(f_{ab}{}^{c})
~=~\epsilon_{a}+\epsilon_{b}+\epsilon_{c}
~=~ \epsilon(\gamma_{abc})+1~=~ \epsilon({a}_{bac})+1~.
\eeq
With the help of the identity
\beq
(e^{T})_{a}{}^{A}~(\larrow{\partial^{l}_{A}}g_{CB})~
e^{B}{}_{b}~e^{C}{}_{c}(-1)^{\epsilon_{b}\epsilon_{C}} 
~=~\gamma_{abc}+(-1)^{\epsilon_{b}\epsilon_{c}}\gamma_{acb}
\equiv\gamma_{a\{bc\}} ~,
\eeq
the Levi-Civita formula \eq{famlevicivita} translates into
\bea
2 \Gamma_{c,ab}
&=&(-1)^{\epsilon_{c}\epsilon_{a}}\gamma_{a\{cb\}}
+(-1)^{\epsilon_{b}(\epsilon_{a}+\epsilon_{c})}\gamma_{b\{ca\}}
-\gamma_{c\{ab\}}  \cr
&=&(-1)^{\epsilon_{c}\epsilon_{a}}\gamma_{acb}
+(-1)^{\epsilon_{c}(\epsilon_{a}+\epsilon_{b})}\gamma_{abc} \cr
&&+(-1)^{\epsilon_{a}\epsilon_{b}+\epsilon_{a}\epsilon_{c}
+\epsilon_{b}\epsilon_{c}}\gamma_{bac}
+(-1)^{\epsilon_{b}(\epsilon_{a}+\epsilon_{c})}\gamma_{bca} \cr
&&-\gamma_{cab}-(-1)^{\epsilon_{a}\epsilon_{b}}\gamma_{cba} ~.  
\eea
If $M$ is an odd K\"{a}hler manifold wrt.\ the canonical $P$-structure,
the only non-vanishing components of the raised Christoffel symbols 
\mb{\Gamma^{b}{}_{ac}} are the pure bosonic and the pure fermionic 
components. {\em We assume for the rest of this paper the symplectic
condition \eq{reallycove}.}

\subsection{Jacobi Identity}
\label{secjac}

\noi
Certain (partially) antisymmetrized versions of the above mentioned 
structure functions are independent of the curved coordinate system 
that was used to define them in the first place. We mention, most notable
\beq
 f_{[ab]}{}^{c}~=~ f_{ab}{}^{c}
-(-1)^{\epsilon_{a}\epsilon_{b}} f_{ba}{}^{c}~,~~~~~~~~~~
{a}_{b[ac]}~=~{a}_{bac}-(-1)^{\epsilon_{a}\epsilon_{c}} {a}_{bca}~.
\eeq 
The Jacobi identity for the anti-bracket, or equivalently the closeness
of $E$ yields Jacobi identifies for the structure functions
\bea
  \sum_{{\rm cycl.}~a,b,c} (-1)^{\epsilon_{a}\epsilon_{c}}
f_{[ab]}{}^{d}~E^{(0)}_{dc}  &=&0~, \cr
 \sum_{{\rm cycl.}~a,b,c} 
 (-1)^{\epsilon_{b}(\epsilon_{c}+1)} a_{b[ac]} &=&0~.
\label{fjacobi}
\eea
The differential operators \mb{\larrow{\partial^{l(\Gamm)}_{a}}
\equiv (e^{T})_{a}{}^{A}\larrow{\partial^{l}_{A}}} {}form an algebra
\beq
 [\larrow{\partial^{l(\Gamm)}_{a}}, \larrow{\partial^{l(\Gamm)}_{b}}]
   ~=~f_{[ab]}{}^{c} \larrow{\partial^{l(\Gamm)}_{c}}~.
\label{falgebra}
\eeq
The corresponding Jacobi identity is
\beq
   \sum_{{\rm cycl.}~a,b,c} (-1)^{\epsilon_{a}\epsilon_{c}}\left(
(\larrow{\partial^{l(\Gamm)}_{a}} f_{[bc]}{}^{e})
-f_{[ab]}{}^{d}f_{[dc]}{}^{e} \right)~=~0~.
\eeq

\vs
\noi
In case of $P$-adapted curved coordinates in an odd K\"{a}hler manifold, 
the structure functions \mb{f_{ab}{}^{c}} has only ``holomorphic'' and 
``anti-holomorphic'' components. To see this, first note that this is true 
independent of the coordinate system for the anti-symmetrized structure 
functions \mb{f_{[ab]}{}^{c}}. This yields that the algebra of differential 
operators \mb{\larrow{\partial^{l(\Gamm)}_{a}}
\equiv (e^{T})_{a}{}^{A}\larrow{\partial^{l}_{A}}} reduces to a 
``holomorphic'' and an ``anti-holomorphic'' algebra, that are mutually 
commutative, cf.\ \eq{falgebra}. For $P$-adapted curved coordinates, 
it follows from the very definition of the structure functions, 
cf.\ \eq{verydefoff}, that
\bea
\epsilon_{b}~\neq~\epsilon_{c}
 &~ \Rightarrow~ & f_{a}{}^{b}{}_{c}~=0~=~f_{ab}{}^{c}~, \cr
\epsilon_{b}~=~\epsilon_{c}
 &~ \Rightarrow~ &\gamma_{abc} ~=0~=~{a}_{bac}~.
\eea
We obtain the claim by combining these arguments.

\subsection{Spin Connection}

\noi
Let us introduce a spin connection 
\mb{\nabla^{\AAA}=d+\AAA~:~TM \times TW \to TW},
where
\beq
\AAA~=~d\Gamm^{A}~ \partial^{r}_{b}~\AAA^{b}{}_{Ac}~\larrow{d\www^{c}} 
 ~=~e^{a}~ \partial^{r}_{b}~\AAA^{b}{}_{ac}~\larrow{d\www^{c}} ~, 
\eeq
and where
\beq
   \AAA^{b}{}_{Ac}~=~ (-1)^{(\epsilon_{A}+\epsilon_{a})\epsilon_{b}} 
 (e^T)_{A}{}^{a}~\AAA^{b}{}_{ac}~.
\eeq
The gauge transformations reads
\beq
  (-1)^{\epsilon_{A}\epsilon_{b}} \AAA^{b}{}_{Ac}~=~
(\larrow{\partial^{l}_{A}}(\Lambda^{T})_{c}{}^{d})(\Lambda^{-1,T})_{d}{}^{b}
(-1)^{\epsilon_{c}(\epsilon_{b}+1)}
+ (-1)^{\epsilon_{A}\epsilon_{b}}(\Lambda^{-1})^{b}{}_{d} \AAA'^{d}{}_{Af}
\Lambda^{f}{}_{c}~.
\eeq
It is assumed that $A$ does 
not depend on the choice of the curved coordinate system.
Of course, the full connection is \mb{\nabla=d+\Gamma+A},
where $\Gamma$ is the Levi-Civita Christoffel symbols.
The spin connection acts trivially on objects 
that carry no flat indices. 
The torsion two-form 
\beq
   T~\in~\Gamma(TW \otimes \Lambda^{2}(T^{*}M))
\eeq
is by definition
\beq
 \hf d\Gamm^{A} \wedge \partial^{r}_{b}~ T^{b}{}_{AC}~ d\Gamm^{C}
~=~ \hf e^{a} \wedge \partial^{r}_{b}~ T^{b}{}_{ac}~ e^{c}
~=~ T~=~ [\nabla \stackrel{\wedge}{,} e]~.
\eeq
$T$ only depends on $\AAA$, because $\Gamma$ has no torsion, cf.\ 
 \eq{notorsion}. In fact, we have
\beq
 T(X,Y)~=~\nabla^{\AAA}_{X}eY-(-1)^{\epsilon(X)\epsilon(Y)}
\nabla^{\AAA}_{Y}eX-e[X,Y]
~=~-(-1)^{\epsilon(X)\epsilon(Y)}T(Y,X)~.
\eeq
In components
\bea
(-1)^{\epsilon_{A}}\larrow{d\www^{b}}
(T(\partial^{r}_{A},\partial^{r}_{C}))
~=~ T^{b}{}_{AC}&=&(-1)^{\epsilon_{b}\epsilon_{A}}
(\larrow{\partial^{l}_{A}}e^{b}{}_{C})
  + ~\AAA^{b}{}_{Ac}~e^{c}{}_{C} \cr
&& +(-1)^{(\epsilon_{A}+1)(\epsilon_{C}+1)}(A \leftrightarrow C)~,
\eea
or
\beq
 T^{b}{}_{ac}~=~(-1)^{\epsilon_{b}\epsilon_{a}}(e^T)_{a}{}^{A}
   (\larrow{\partial^{l}_{A}}e^{b}{}_{C}) e^{C}{}_{c}
  + ~\AAA^{b}{}_{ac}
 +(-1)^{(\epsilon_{a}+1)(\epsilon_{c}+1)}(a \leftrightarrow c)~,
\eeq
It is convenient to lower the first index with the flat metric 
\mb{g^{(0)}_{ab}}:
\bea
  \AAA_{bac}&=&(-1)^{\epsilon_{a}}~ g^{(0)}_{bd}~\AAA^{d}{}_{ac}~, \cr
  T_{bac}&=&(-1)^{\epsilon_{a}}~ g^{(0)}_{bd}~T^{d}{}_{ac}~.
\eea
We then have
\beq
 T_{bac}~=~\aaa_{bac}+\AAA_{bac}
-(-1)^{\epsilon_{a}\epsilon_{c}}(a \leftrightarrow c)
~=~\aaa_{b[ac]}+\AAA_{b[ac]}~.
\eeq

\subsection{Curvature}

\noi
The curvature is
\beq
  R~=~\hf [\nabla \stackrel{\wedge}{,}\nabla]
~=~\hf [\nabla^{\Gamma} \stackrel{\wedge}{,}\nabla^{\Gamma}]
+\hf [\nabla^{\AAA} \stackrel{\wedge}{,}\nabla^{\AAA}]
~=~R^{\Gamma}+R^{\AAA} ~.
\eeq
The Bianchi identity is a trivial consequence of the supercommutator 
version of the Jacobi identity:
\beq
 [\nabla \stackrel{\wedge}{,}R]~=~0~.
\eeq
The curvature of the spin connection is
\beq
 R^{\AAA} ~=~ - e^{d} \wedge e^{a}\otimes\left( 
(-1)^{\epsilon_{a}\epsilon_{b}}\partial^{r}_{b}~ 
(\larrow{\partial^{l(z)}_{a}}\AAA^{b}{}_{df})
+ \partial^{r}_{b} ~\AAA^{b}{}_{ac}\AAA^{c}{}_{df}
- f_{ad}{}^{c}~\partial^{r}_{b} ~\AAA^{b}{}_{cf} 
\right)\otimes d\www^{f}~,
\eeq
The Ricci two-form $\RR^{\AAA}$ thus reads
\beq
 \RR^{\AAA} ~=~ - e^{b} \wedge e^{a}\otimes\left( 
(\larrow{\partial^{l(z)}_{a}}\AAA_{b}) - f_{ab}{}^{c}~\AAA_{c} 
\right)\otimes d\www^{f}~,
\eeq
where the superdeterminant gauge field reads
\beq
  \AAA_{a}~=~(-1)^{(\epsilon_{a}+1)\epsilon_{b}}\AAA^{b}{}_{ab}~.
\eeq 
It transforms under gauge transformations
\mb{\Lambda^{a}{}_{b}} as a tensor 
\beq
  \AAA_{a}~=~(\Lambda^{T})^{a}{}_{b}~\AAA_{b}' ~
 + (\larrow{ \partial^{l(z)}_{a}} \ln {\rm sdet}
(\Lambda^{\cdot}{}_{\cdot} ) )~=~ (\Lambda^{T})^{a}{}_{b}~\AAA_{b}' ~,
\eeq
because \mb{{\rm sdet}(\Lambda^{\cdot}{}_{\cdot})=1}.
The condition for Ricci-form-flatness reads
\beq
  (\larrow{\partial^{l(z)}_{[a}}\AAA_{b]}) - f_{[ab]}{}^{c}~\AAA_{c} ~=~0~. 
\eeq
This is the closeness-condition in a non-Abelian basis that by the 
Poincar\'{e} Lemma ensures that $\AAA_{a}$ is locally exact.
We may also form a determinant gauge field
\beq
  \tilde{\AAA}_{a}~=~(-1)^{\epsilon_{a}\epsilon_{b}}\AAA^{b}{}_{ab}~.
\eeq 
It also transforms under gauge transformations
\mb{\Lambda^{a}{}_{b}} as a tensor 
\beq
  \tilde{\AAA}_{a}~=~(\Lambda^{T})^{a}{}_{b}~\tilde{\AAA}_{b}' ~
 + (\larrow{ \partial^{l(z)}_{a}}  \ln {\rm det}
(\Lambda^{\cdot}{}_{\cdot} ) )
~=~(\Lambda^{T})^{a}{}_{b}~\tilde{\AAA}_{b}' ~,
\eeq
because \mb{ {\rm tr}((\ln\Lambda)^{\cdot}{}_{\cdot})=
 \ln {\rm det}(\Lambda^{\cdot}{}_{\cdot})=0}.
Here we made use of the fact that the matrices $\Lambda$ have 
vanishing bose-fermi blocks.

\subsection{Levi-Civita Spin Connection}

\noi
As in the usual bosonic case we define the
Levi-Civita spin connection to be the unique spin connection that
satisfies
\beq
  T~=~0 ~,~~~~~~~~~~~~~~~~~~~~\nabla g^{(0)}~=~0~.
\label{levicivitaspin}
\eeq
The last equation yields 
\beq
 \AAA_{bac}+(-1)^{\epsilon_{b}\epsilon_{c}
+\epsilon_{a}(\epsilon_{b}+\epsilon_{c})}\AAA_{cab} ~=~0~.
\eeq
Together with the condition of no torsion, this implies that
the \mb{\AAA_{bac}} can be expressed purely in terms of the structure
functions \mb{\aaa_{b[ac]}}:
\bea
 2 \AAA_{bac}&=&-(-1)^{\epsilon_{a}\epsilon_{b}}\aaa_{a[bc]}
-(-1)^{\epsilon_{c}(\epsilon_{a}+\epsilon_{b})} \aaa_{c[ba]}
-\aaa_{b[ac]} \cr
 &=&(-1)^{\epsilon_{b}(\epsilon_{a}+\epsilon_{c})}\aaa_{acb}
-(-1)^{\epsilon_{a}\epsilon_{b}}\aaa_{abc} \cr
&&+ (-1)^{\epsilon_{a}\epsilon_{b}+\epsilon_{a}\epsilon_{c}
+\epsilon_{b}\epsilon_{c}}\aaa_{cab}
-(-1)^{\epsilon_{c}(\epsilon_{a}+\epsilon_{b})} \aaa_{cba} \cr
&&+(-1)^{\epsilon_{a}\epsilon_{c}}\aaa_{bca}-\aaa_{bac}~.
\eea
Evidently, \mb{\AAA_{bac}} does not tranform under change of the
curved coordinate frame.
On the other hand we can construct a spin connection by conjugating with
the vielbein
\beq
 \nabla_{X}^{\AAA}~=~e~ \nabla_{X}^{\Gamma}~e^{-1}~.
\label{conjconj}
\eeq
It satisfies \eq{levicivitaspin}, so it is the Levi-Civita spin connection.
It follows that
\bea 
\Gamma_{b,ac}&=&\AAA_{bac}+\aaa_{bac}~, \cr
   \AAA^{b}{}_{ac}-\Gamma^{b}{}_{ac}
&=&(-1)^{\epsilon_{b}(\epsilon_{a}+\epsilon_{c})+\epsilon_{c}}f_{ac}{}^{b}
~=~-(-1)^{\epsilon_{b}\epsilon_{a}}f_{a}{}^{b}{}_{c}~.
\label{eurika}
\eea
Conjugation with vielbeins does not change the curvature:
\beq
  R^{\AAA}~\cong~R^{\Gamma}~.
\eeq
Here we have identified the $w$-basis with the vielbein-basis.

\vs
\noi
From the conjugation formula \eq{conjconj} it follows that
the Levi-Civita spin connection respects the flat anti-symplectic 
structure \mb{\nabla E^{(0)}=0} and the flat canonical almost parity 
structure \mb{\nabla P^{(0)}=0}, where 
\mb{P^{(0)a}{}_{b}} \mb{=} \mb{(-1)^{\epsilon_{a}} \delta^{a}_{b}}.
Explicitly, the condition \mb{\nabla P^{(0)}=0} reads
\beq
    \epsilon_{b}=\epsilon_{c}~~\Rightarrow~~A_{bac}~=~0~.
\eeq
This condition is of course completely superseded,
if we furthermore assume that the manifold $M$ is an odd 
K\"{a}hler manifold wrt.\ the canonical $P$-structure.  Then 
only the bose-bose-bose and the fermi-fermi-fermi components of
\mb{\AAA^{b}{}_{ac}} survive. This follows from arguments similar
to those presented in Section~\ref{conparity} for a tangent bundle
connection \mb{\nabla^{\Gamma}}, which yields the analogous conclusion
for the Christoffel symbols \mb{\Gamma^{b}{}_{ac}}.
Or one could use the relation \eq{eurika} and the fact that the structure 
functions \mb{f_{a}{}^{b}{}_{c}} and \mb{f_{ab}{}^{c}}
has also only ``holomorphic'' and ``anti-holomorphic'' components.

\subsection{Connection Fields}

\noi
We translate the superdeterminant connection field $F_{A}$ into 
the flat indices as follows:
\beq
  (-1)^{\epsilon_{a}}F_{a}~=~\Gamma^{b}{}_{ba}
~=~ (-1)^{\epsilon_{A}}F_{A}~e^{A}{}_{a}
~=~(-1)^{\epsilon_{b}} f_{b}{}^{b}{}_{a}+ (-1)^{\epsilon_{a}}\fo_{a}~.
\eeq
Here \mb{\fo_{a}} is
\bea
(-1)^{\epsilon_{a}}\fo_{a}&=&\AAA^{b}{}_{ba}
~=~E_{(0)}^{dc}~f_{cd}{}^{b}~g^{(0)}_{ba} \cr
&=& \hf (-1)^{\epsilon_{d}} g_{(0)}^{dc}~f_{[cd]}{}^{b}~g^{(0)}_{ba} 
~=~  - f_{[ab]}{}^{b}  ~.
\eea
The last equality follows from the Jacobi identity \eq{fjacobi}.
{}From \eq{eurika} follows a relation between the two 
superdeterminant connection fields
\beq
  F_{a}-\AAA_{a}~=~ (-1)^{\epsilon_{b}}f_{a}{}^{b}{}_{b}
~=~(-1)^{\epsilon_{b}} (\larrow{\partial^{l(z)}_{a}} e^{b}{}_{B}) e^{B}{}_{b} 
~=~(\larrow{\partial^{l(z)}_{a}} \ln {\rm sdet}(e))~.
\eeq
So for the Levi-Civita connection we have that
\beq
   F_{a} ~=~ (\larrow{\partial^{l(z)}_{a}} \ln {\rm sdet}(e))
~~\Leftrightarrow~~ \AAA_{a}~=~0~.
\eeq
We have that the quantity \mb{F^{(0)}_{A}} defined in \eq{alsonice} 
transforms into
\bea
   F^{(0)}_{a}&=&(e^{T})_{a}{}^{A}~F^{(0)}_{A}
~=~-E_{ab}~\Gamma^{b}{}_{cd}~E^{dc} \cr
&=&-(-1)^{\epsilon_{a}}f_{[ab]}{}^{b}+\tilde{\fo}_{a}
~=~\fo_{a}+\tilde{\fo}_{a} ~.
\eea
Here \mb{\tilde{\fo}_{a}} is 
\beq
\tilde{\fo}_{a}~=~g_{(0)}^{dc}~f_{cd}{}^{b}~g^{(0)}_{ba} 
~=~ \hf (-1)^{\epsilon_{d}} E_{(0)}^{dc}~f_{\{cd\}}{}^{b}~g^{(0)}_{ba}~.
\eeq
The connection field \mb{\tilde{F}_{A}} is
\beq
  \tilde{F}_{A}~=~(\larrow{\partial^{l}_{A}}e^{b}{}_{B})~e^{B}{}_{b}
~=~(e^{T})_{A}{}^{a} f_{a}{}^{b}{}_{b} ~.
\eeq
In the flat indices it reads 
\beq
  \tilde{F}_{a}~=~(-1)^{\epsilon_{a}+\epsilon_{b}}\Gamma^{b}{}_{ba}
~=~ (e^{T})_{a}{}^{A}~ \tilde{F}_{A}
~=~ f_{a}{}^{b}{}_{b}~.
\eeq
{}From \eq{eurika} follows a relation between the two 
determinant connection fields
\beq
  \tilde{F}_{a}-\tilde{\AAA}_{a}~=~ f_{a}{}^{b}{}_{b}
~=~ (\larrow{\partial^{l(z)}_{a}}e^{b}{}_{B}) e^{B}{}_{b} 
~=~(\larrow{\partial^{l(z)}_{a}} \ln\tilde{\rho} )~.
\eeq
This is in agreement with the fact that \mb{\tilde{\AAA}_{a}=0}
for the Levi-Civita spin connection.
The determinant density is
\beq
    \ln\tilde{\rho}~=~(\ln e)^{a}{}_{a}~.
\eeq
The odd Laplacian in flat indices reads
\beq
  \Delta~=~\hf g_{(0)}^{ab}~(\fo_{b}+ \larrow{\partial^{l(\Gamm)}_{b}})
 \larrow{\partial^{l(\Gamm)}_{a}} 
~=~~:\hf g_{(0)}^{ab}~\larrow{\partial^{l(\Gamm)}_{b}}
 \larrow{\partial^{l(\Gamm)}_{a}} :~
+~\hf g_{(0)}^{ab}~ F^{(0)}_{b}
\larrow{\partial^{l(\Gamm)}_{a}}  ~.
\eeq
We note the identity
\beq
   (-1)^{\epsilon_{A}} (\larrow{\partial^{l}_{A}}e^{A}{}_{a})
~=~-(-1)^{\epsilon_{b}} f_{b}{}^{b}{}_{a}~.
\eeq

\subsection{Odd K\"{a}hler Manifolds}

\noi
Finally, let us just list the case of an odd K\"{a}hler manifold $M$. Then
\beq
  \Delta~=~~:\hf g_{(0)}^{ab}~\larrow{\partial^{l(\Gamm)}_{b}}
 \larrow{\partial^{l(\Gamm)}_{a}} :
~~=~\hf g_{(0)}^{ab}~\larrow{\partial^{l(\Gamm)}_{b}}
 \larrow{\partial^{l(\Gamm)}_{a}} ~,
\eeq
\beq
  F^{(0)}_{a}~=~0~,~~~~~~ \fo_{a}~=~0~,~~~~~~\tilde{\fo}_{a}~=~0~, 
\eeq
The following formulas apply to the non-vanishing components only:
\bea
g^{(0)}_{ab}&=&(~:~\larrow{\partial^{l(\Gamm)}_{a}}
\larrow{\partial^{l(\Gamm)}_{b}}~:~K)~,\cr
\Gamma_{c,ab}&=&(K~:~\rarrow{\partial^{r(z)}_{c}}
\rarrow{\partial^{r(z)}_{a}}\rarrow{\partial^{r(z)}_{b}}~:~  )~,   
\eea
where $K$ is the odd K\"{a}hler potential.
In case of a Ricci-form-flat manifold $M$ we have
\beq
  \AAA_{a}~=~0~,~~~~~~~~~
F_{a}~=~(\larrow{\partial^{l(z)}_{a}} \ln\rho )~,~~~~~~~
\rho~=~{\rm sdet}(e)~.
\eeq

\setcounter{equation}{0}
\begin{appendix}
\section{Super conventions}

\noi
Derivatives have two kinds of attributes.
First, a superscript ``$r$'' or ``$l$'' indicates a left or right
derivative
\beq
{{\partial^{l}} \over {\partial {\Gamm^{A}}}}
~=~(-1)^{\epsilon_{A}}{{\partial^{r}} \over {\partial {\Gamm^{A}}}}~.
\eeq
Secondly, arrows indicate, on which objects the derivatives should act.
The derivatives is uniquely specified through its action on the fundamental 
variables
\beq
(\papal{\Gamm^{A}}\Gamm^{B})~=~\delta^{B}_{A}
~=~(\Gamm^{B}\papar{\Gamm^{A}})~.
\eeq
If a derivative carries no arrow it does not act on anything
present in the formula.
It is then merely understood as the natural basis for the 
tangent space in the sense that we are only interested in immitating 
its transformation properties under change of coordinates.
\beq
 {{\partial^{r}} \over {\partial {\Gamm^{A}}}} 
~=~{{\partial^{r}} \over {\partial {\Gamm'^{B}}}}~ 
(\Gamm'^{B}\papar{\Gamm^{A}})
~=~(-1)^{\epsilon_{A}+\epsilon_{B}} (\papal{\Gamm^{A}}\Gamm'^{B})~
{{\partial^{r}} \over {\partial {\Gamm'^{B}}}}
\eeq
and
\beq
 {{\partial^{l}} \over {\partial {\Gamm^{A}}}} 
~=~(\papal{\Gamm^{A}}\Gamm'^{B})~
{{\partial^{l}} \over {\partial {\Gamm'^{B}}}}~. 
\eeq
The exterior derivative $d$ is
\beq
 d~=~d\Gamm^{A} \papal{\Gamm^{A}}~.
\eeq
One-forms act on vectors according to
\beq
(-1)^{\epsilon_{B}}\larrow{d\Gamm^{A}}(\partial^{r}_{B})~=~
\larrow{d\Gamm^{A}}(\partial^{l}_{B})
~=~(-1)^{\epsilon_{A}\epsilon_{B}} (\papal{\Gamm^{B}}\Gamm^{A})
~=~(-1)^{\epsilon_{A}}\delta^{A}_{B}~.
\eeq
Note that this definition is stabile under change of coordinates as it 
should be. The contraction $i_{X}$ with a bosonic vector field $X$, 
\mb{\epsilon(X)=0},  
\beq
i_{X}~=~X^{A}~ i^{l}_{A}~=~ i^{r}_{A}~X^{A} 
\eeq 
is defined via its action on the natural basis of one-forms
\beq
(-1)^{\epsilon_{A}}\larrow{i^{r}_{A}}(d\Gamm^{B})
~=~\larrow{i^{l}_{A}}(d\Gamm^{B})~=~\delta^{A}_{B}~.
\eeq
In fact, we can symbolically write
\beq
\larrow{i^{l}_{A}}~=~ \papal{(d\Gamm^{A})}  ~.
\eeq
Both the exterior derivative $d$ and the contraction $i_{X}$ carries
odd ``form-parity''
\beq
p(d)~=~1~=~p(i_{X})~.
\eeq
The wedge $\wedge$ will in this respect be a total redundant symbol,
\ie for two forms $\omega$ and $ \eta$ 
\beq
\omega \wedge \eta ~=~
(-1)^{\epsilon(\omega)\epsilon(\eta)+p(\omega)p(\eta)}\eta \wedge\omega~.
\eeq
In the same spirit we define the supercommutator $[A,B]$ of two operators 
$A$ and $B$ to be
\beq
 [A,B]~=~AB-(-1)^{\epsilon(A)\epsilon(B)+p(A)p(B)}BA
\eeq
For instance
\beq
 d^2~=~\hf[d,d]~=~0~,~~~~
[i_{X},i_{Y}]~=~0~,~~~~~{\cal L}_{X}~=~[i_{X},d]~.
\eeq
\end{appendix}



\noi
{\bf Acknowledgements}.
I would like to thank A.~Nersessian, O.M.~Khurdaverdian,
A.~Schwarz and B.~Zwiebach for discussions. The research is supported
by the Danish Natural Science Research Council(SNF), grant no.~9602107.


\end{document}